\documentclass[12pt]{article}
\usepackage{UF_FRED_paper_style}
\usepackage{times}
\usepackage{amsmath, amssymb}
\usepackage{geometry}
\usepackage{cancel}
\usepackage{marginnote}
\usepackage{amsthm}
\usepackage{cleveref}
\usepackage{natbib}
\usepackage{appendix}
\usepackage{booktabs}

\hypersetup{
    colorlinks = true,
    citecolor = blue
}
\usepackage{lipsum}  
\usepackage{setspace}

\usepackage{bm}
\usepackage{booktabs}
\usepackage{subcaption}
\usepackage{graphicx}
\title{\textbf{Power and Sample Size Calculations for Bayes Factors in two-arm clinical Phase II Trials with binary Endpoints}}

\author{Riko Kelter\thanks{Correspondence concerning this article should be addressed to \url{rkelter@uni-koeln.de}.
    Draft version 1.0, 23/12/25. This paper has not been peer reviewed. Please do not copy or cite without author's permission. Data and R code to reproduce our results are openly available at \url{https://osf.io/zsrfh/overview?view_only=351aecc896d0468991569c99ae58beaf}. The R package \texttt{bfbin2arm} is available on CRAN, see \url{https://cran.r-project.org/web/packages/bfbin2arm/index.html}. We declare no conflict of interest.}\\
	Institute of Medical Statistics and Computational Biology\\
	Faculty of Medicine\\
    University of Cologne\\
    Cologne, Germany
    }

\date{\today}

\begin{document}

{\setstretch{.8}
\maketitle
\begin{abstract}
Bayesian sample size calculations in clinical trials usually rely on complex Monte Carlo simulations in practice. Obtaining bounds on Bayesian notions of the false-positive rate and power often lack closed-form or approximate numerical solutions. In this paper, we focus on power and sample size calculations for Bayes factors in the two-arm binomial setting of phase II trials. We cover point-null versus composite and directional hypothesis tests, derive the corresponding Bayes factors, and discuss relevant aspects to consider when pursuing Bayesian design of experiments with the introduced approach. Based on these Bayes factors, we propose a numerical approach which allows to determine the necessary sample size to obtain prespecified bounds of Bayesian power and type-I-error rate in a computationally efficient way. Our method does not rely on Monte Carlo simulations and instead solely relies on standard numerical methods. Real-world examples of phase II trials from oncology and autoimmune diseases illustrate the advantage of the proposed calibration method. In summary, our approach allows for a Bayes-frequentist compromise by providing a Bayesian analogue to a frequentist power analysis for various Bayes factors in the two-arm binomial setting of a phase II clinical trial. The methods are implemented in our R package \texttt{bfbin2arm}.


\noindent

\end{abstract}
}

\section{Introduction}\label{sec:intro}

Sample size determination is one of the most critical aspects of clinical trial design, ensuring that studies are both scientifically informative and ethically justified. Inadequate sample sizes can lead to inconclusive results, while excessive recruitment wastes resources and may unnecessarily expose patients to experimental treatments \citep{Grieve2022,Spiegelhalter2004,Chow2008}. This balance between statistical rigor and ethical responsibility is particularly pronounced in early-phase clinical trials, where therapeutic uncertainty is large and patient populations are often small. 

In phase II oncology trials, where the primary endpoint is often binary (e.g., tumor response, remission, or toxicity), classical methods for design such as Simon’s two-stage procedure \citep{Simon1989} remain widely used. These frequentist approaches allow explicit control of type-I and type-II error rates, typically set at $\alpha=0.05$ and $\beta=0.2$, respectively. However, they rely on fixed-sample inference and do not naturally incorporate prior information about treatment effect—a limitation increasingly recognized in modern adaptive and confirmatory settings \citep{Berry2006,Spiegelhalter2004,Thall1994}. 

Bayesian designs provide a principled and flexible alternative. They formally integrate prior evidence from previous studies, expert opinion, or mechanistic modeling into the analysis, updating this knowledge as data accumulate. In Bayesian phase II designs, decision rules are typically expressed through posterior probabilities or Bayes factors. For example, investigators may recommend advancing a treatment to phase III if the posterior probability that the response rate exceeds a clinically relevant threshold surpasses a prespecified decision boundary \citep{Thall2007,Thall1994,Neuenschwander2009,Grieve2016}. Bayesian designs also allow adaptive stopping for futility or efficacy while maintaining interpretable long-run error control \citep{Berry2006}.

The Bayes factor, in particular, provides an appealing index of evidence for or against a null hypothesis by comparing marginal likelihoods:
\begin{align}\label{eq:bayesFactorIntro}
   \underbrace{\frac{P(H_0 \mid y)}{P(H_1 \mid y)}}_{\text{Posterior odds}} = 
   \underbrace{\frac{f(y \mid H_0)}{f(y \mid H_1)}}_{\text{Bayes factor $\mathrm{BF}_{01}(y)$}} \cdot 
   \underbrace{\frac{P(H_0)}{P(H_1)}}_{\text{Prior odds}},
\end{align}
Here, $\mathrm{BF}_{01}(y)$ quantifies how much more likely the observed data are under $H_0$ than $H_1$ \citep{KassRaftery1995}. For example, $\mathrm{BF}_{01}(y)=10$ indicates that the data support $H_0$ ten times more strongly than $H_1$. Because the Bayes factor is independent of prior odds assigned to the hypotheses, it is often considered a more objective Bayesian evidence measure than posterior probabilities. It has become increasingly popular in biomedical research as a bridge between Bayesian and frequentist reasoning \citep{Rouder2009,Pourmohammad2023,PawelHeld2024,KelterPawel2025,Bartos2022}.

A major challenge, however, lies in Bayesian sample size planning. Unlike classical power and sample size formulas derived from asymptotic theory, Bayesian designs rarely yield closed-form expressions for required sample sizes. In most cases, power and type-I-error rates must be evaluated via simulation, particularly when the prior distribution or decision threshold introduces nonlinear dependencies on the parameter of interest. Yet, using frequentist summaries of Bayesian criteria -- so-called ``calibrated Bayes'' methods \citep{Dawid1982,rubinBayesianlyJustifiableRelevant1984,Little2006,Grieve2016} -- provides a compelling compromise. Regulatory authorities such as the U.S. Food and Drug Administration (FDA) or the European Medicines Agency (EMA) likewise encourage Bayesian analyses that demonstrate acceptable frequentist operating characteristics \citep{FDABayes2010,ema2022reflectionpaper,ionanBayesianMethodsHuman2023}. 

This hybrid approach involves quantifying the long-run performance of Bayesian decision metrics such as the Bayes factor or posterior probability. For a Bayes factor $\mathrm{BF}_{01}(y)$ in favor of $H_0$ and a decision threshold $k<1$, the Bayesian analogues of a type-I-error rate and power are given by
\begin{align}\label{eq:bayesianPower}
   P(\mathrm{BF}_{01}(y) < k \mid H_0) \quad \text{and} \quad P(\mathrm{BF}_{01}(y) < k \mid H_1),
\end{align}
respectively. The probability $P(\mathrm{BF}_{01}(y) < k \mid H_0)$ can be interpreted as the chance under $H_0$ to find at least evidence $1/k$ in favour of $H_1$ -- which resembles a false-positive decision -- and thus constitutes a Bayesian type-I-error rate.\footnote{This holds, because $\mathrm{BF}_{01}(y)=1/\mathrm{BF}_{10}(y)$ and thus, $\mathrm{BF}_{01}(y) < k \Leftrightarrow \mathrm{BF}_{10}(y) \geq 1/k$.} The probability $P(\mathrm{BF}_{01}(y) < k \mid H_1)$ constitutes a notion of Bayesian power based on similar arguments.

Using Jeffreys’ scale \citep{Jeffreys1939}, $k=1/10$ corresponds to “strong” evidence against $H_0$. If $P(\mathrm{BF}_{01}(y)<1/10 \mid H_0)\leq \alpha$ and $P(\mathrm{BF}_{01}(y)<1/10 \mid H_1)\geq 1-\beta$, then the Bayesian test achieves type-I-error control at level $\alpha$ and power $1-\beta$:
\begin{align}\label{eq:bayesianDesign}
   P(\mathrm{BF}_{01}(y) < k \mid H_0)\leq \alpha \quad \text{and} \quad P(\mathrm{BF}_{01}(y) < k \mid H_1)> 1-\beta.
\end{align}
These quantities naturally align with the control of false positive and false negative decisions in a frequentist sense while maintaining a fundamentally Bayesian interpretation.

In the context of binary outcomes, such as tumor response rates in oncology studies, sample size planning under \Cref{eq:bayesianDesign} offers several practical and conceptual advantages. The binomial likelihood provides a natural setting for analytical treatment through conjugate Beta priors, while the Bayes factor facilitates transparent assessment of evidence without relying on asymptotic approximations \citep{Ibrahim2001,Neuenschwander2009,Spiegelhalter2004}. Moreover, Bayesian predictive power or ``assurance''—the prior-averaged probability of achieving favorable evidence under the design—can be calculated to assess trial robustness under prior uncertainty \citep{Case2022,Grieve2016,Grieve2022}. These properties make Bayesian trial designs particularly useful for adaptive, small-sample, or rare-disease settings \citep{Berry2006}. Henceforth, when we speak of assurance or the \textit{probability of compelling evidence under $H_0$}, we refer to the following:
\begin{align}\label{eq:pce_H0}
    P(\mathrm{BF}_{01}(y)>k_f|H_0)
\end{align}
Requiring $P(\mathrm{BF}_{01}(y)>k_f|H_0)>1-\beta_f$ for some $\beta_f$ implies that when $H_0$ holds, there is a minimum assured probability of finding at least evidence $k_f$ in favour of $H_0$. This undermines trust in the result of the trial, as not only is the Bayesian power under $H_1$ in \Cref{eq:bayesianPower} taken into consideration, but also the probability to find sufficient evidence in favour of $H_0$ if the latter holds. Phrased differently, if there is no effect, a calibrated probability of compelling evidence for $H_0$ -- that is, $P(\mathrm{BF}_{01}(y)>k_f|H_0)>1-\beta_f$ -- asserts that the trial has an adequate chance to reveal this. 

In this paper, we adopt the $\mathrm{BF}_{01}$ orientation of \cite{Jeffreys1939}, with decision thresholds $k<1$ such that small Bayes factors indicate evidence against the null hypothesis. Our focus is to determine the smallest sample size which satisfies the evidence-based calibration criteria in \Cref{eq:bayesianDesign} for given $\alpha$ and $\beta$ levels. We use the $\mathrm{BF}_{01}$ orientation to maintain consistency with Jeffreys’ original formulation and to align the interpretation of ``small values indicating evidence against $H_0$'' with the familiar frequentist convention for $p$-values.

\subsection{Outline}
This article develops a novel approach for Bayesian power and sample size calculation in two-arm (clinical) trials with binary endpoints using Bayes factors. Our approach is simulation-free and uses only standard numerical techniques like numerical integration. This frees users from the necessity to conduct, replicate and report Monte Carlo studies (including the computation and reporting of Monte Carlo standard errors for all estimates, a necessity which is challenging due to several reasons, for details see \cite{Morris2019}, \cite{Siepe2024} or \cite{Kelter2023}). Based on our novel approach, power and sample size calculations for Bayes factors in phase II clinical trials with binary endpoints can be carried out simulation-free, computationally efficient and without relying on asymptotic theory or approximations.

Therefore, \Cref{sec:methodology} introduces the beta-binomial model and explains the general approach in detail. \Cref{subsec:two-sided-hyp} introduces the different hypotheses under consideration in a phase II trial with two trial arms, and provides a full walk-through of our approach for the case of the two-sided hypothesis test, which tests
$$H_0:p_1=p_2 \text{ versus } H_1:p_1 \neq p_2$$
where $p_1$ and $p_2$ are the success probabilities in the control and treatment arm. 

\Cref{subsec:one-sided-hyp1}, \Cref{subsec:one-sided-hyp2} and \Cref{subsec:one-sided-hyp3} cover the relevant derivations and details for the one-sided tests in a phase II trial with two arms and binary endpoints. \Cref{subsec:examples} then provides two-real world examples of phase II trials. The first trial is the riociguat trial investigating the effect of riociguat versus placebo in patients with systemic sclerosis, reported in \citep{khannaRiociguatPatientsEarly2020}. The second trial is the ICT-107 trial, which assessed among other aspects immunologic response in patients with newly diagnosed glioblastoma \citep{wenRandomizedDoubleBlindPlaceboControlled2019}. We showcase the usefulness of our approach by re-analyzing both trials with our novel approach and provide calibrated Bayesian designs using Bayes factors under informative and noninformative design prior choices. Also, we illustrate the effect of different evidence thresholds and provide guidance for practice in applying our method. \Cref{sec:discussion} discusses the results and provides a conclusion of the paper.

\begin{figure}[!htb]
    \centering
    \includegraphics[width=1.0\linewidth]{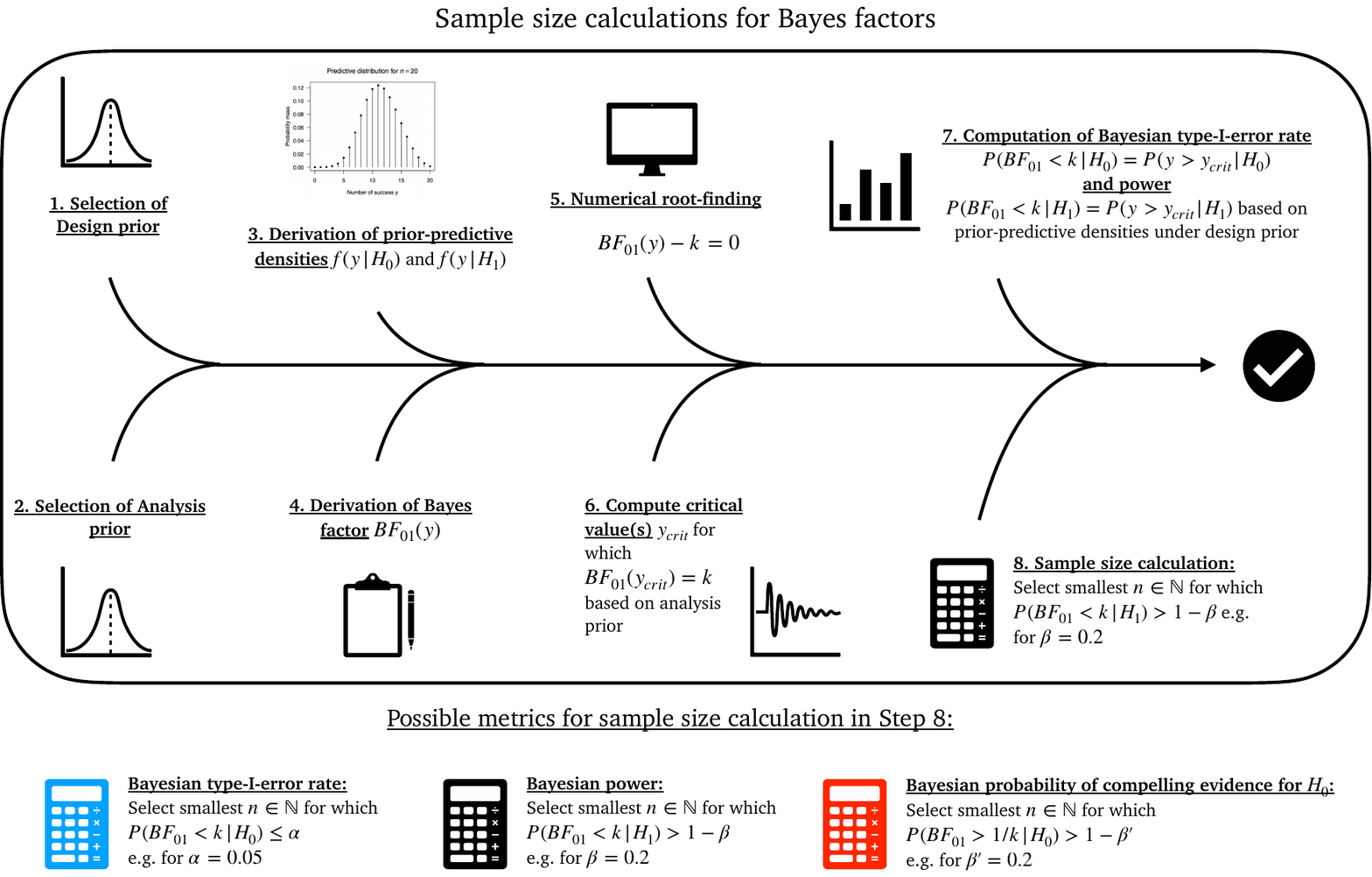}
    \caption{Overview of the methodology underlying Bayesian power and sample size calculations for Bayes factors in the two-arm binomial setting, modified and adapted from the one-arm setting in \cite{KelterPawel2025}}
    \label{fig:overview}
\end{figure}

\section{Bayesian power and sample size calculations for Bayes factors in the two-arm setting}\label{sec:methodology}

\Cref{fig:overview} provides a graphical summary of our approach, modified and adapted from \cite{KelterPawel2025}, where the one-arm case it treated in detail. The idea of our method is to derive the prior-predictive densities $f(y|H_0)$ and $f(y|H_1)$ and the Bayes factor $\mathrm{BF}_{01}(y)$ based on a selected design and analysis prior (steps 1. to 4. in \Cref{fig:overview}). The analysis prior is the prior used for calculation of the Bayes factor, while the design prior is the prior which is employed to calculate the power and sample size a priori before any data are observed at the planning stage of the trial. The next step of our method is to find a numerical solution to the equation
$$\mathrm{BF}_{01}(y)-k=0$$
for a specified evidence threshold $k$ such as $k=1/10$ or $k=1/3$, so that we can compute critical values $y_{crit}$ for which $\mathrm{BF}_{01}(y_{crit})=k$ holds for our analysis prior, and the Bayes factor passes our required threshold for evidence in favour of $H_1$ (steps 5. and 6. in \Cref{fig:overview}). Based on the prior-predictive obtained in the third step, the seventh step then consists in computation of the Bayesian type-I-error rate $P(\mathrm{BF}_{01}(y)<k|H_0)$ and power $P(\mathrm{BF}_{01}(y)<k|H_1)$ based on the prior-predictive densities under the design prior. The last and eigth step of our approach is to select the smallest sample size $n$ for which we achieve the desired power, type-I-error rate or probability of compelling evidence for $H_0$, or a combination of the latter.

\subsection{Two-sided hypothesis test}\label{subsec:two-sided-hyp}
In this section, we illustrate our approach in detail by walking through the process shown in \Cref{fig:overview} for the two-sided hypothesis test $H_0:p_1=p_2$ versus $H_1:p_1\neq p_2$ in a two-arm phase II trial with binary endpoints. Our main assumption here is that the observed data in both groups are from two random variables $Y_1,Y_2$ which both follow a binomial distribution with parameters $n_1$ and $n_2$ and $p_1$ respectively $p_2$,
$$Y_1\sim \mathrm{Bin}(n_1,p_1), \hspace{1cm} Y_2\sim \mathrm{Bin}(n_2,p_2)$$
The probability mass functions are given as
\begin{align}\label{eq:binomialLik}
    &f(y_1 \mid p_1)={n_1\choose y_1}p_1^{y_1} (1-p_1)^{n_1-y_1}, \hspace{1cm} f(y_2 \mid p_2)={n_2\choose y_2}p_2^{y_2} (1-p_2)^{n_2-y_2}
\end{align}
for $y_1=0,...,n_1$ and $y_2=0,...,n_2$. We denote by $Y_1$ the control group data, which reveives the standard of care or possibly a placebo, and by $Y_2$ the treatment group data, which receives the novel drug or treatment under study. In this setting, we have $p_i\in [0,1]$ for $i=1,2$, so the parameter space $\Theta:=[0,1]\times [0,1]$ is the two-dimensional unit cube. The full data vector $D=(Y_1,Y_2)$ has, under the assumption of independence of data in both groups, the following joint density:
\begin{align}\label{eq:jointDensity}
    f(y_1,y_2|p_1,p_2)=\prod_{i=1}^2 p_i^{y_i}(1-p_i)^{n_i-y_i}
\end{align}

Now, in the context of a phase-II-trial two hypotheses could be of interest. First, the test of equality of proportions $p_1=p_2$ in the treatment and control group, where the former receives a novel drug and the latter the standard of care or a placebo. The hypotheses tested are then given as
\begin{align}\label{eq:twoSided}
    H_0:p_1=p_2 \hspace{1cm} \text{ versus } \hspace{1cm} H_1:p_1\neq p_2
\end{align}
Alternatively, a well-known parameterization of this test introduces a difference parameter $\eta=p_2-p_1$ and the grand mean $\zeta=\frac{1}{2}(p_1+p_2)$. Using this parameterization, we have
$$p_1=\zeta-\frac{\eta}{2}, \hspace{1cm} p_2=\zeta+\frac{\eta}{2}$$
and the hypotheses in \Cref{eq:twoSided} can be rewritten as:
\begin{align}\label{eq:twoSidedParameterized}
    H_0:\eta = 0 \hspace{1cm} \text{ versus } \hspace{1cm} H_1:\eta \neq 0
\end{align}
The parameterization was first proposed by \cite{Gunel1974}, see also \cite{Dickey1970a}, \cite{Jamil2017} and \cite{Kelter2025}. Now, in the context of a phase-II-b trial, the test in \Cref{eq:twoSidedParameterized} seems attractive, because it can express evidence for $H_0:\eta=0$. In this case, the standard of care or placebo and the novel treatment are deemed equally effective. If, on the other hand, evidence for $H_1:\eta \neq 0$ is found after analysing the trial data, there are two options. First, $p_1>p_2$, in which case the efficacy in the control group (placebo or standard of care) is even better than in the treatment group. Second, $p_1<p_2$, in which case the treatment works better than the standard of care or placebo in the control group. Both options are possible as a trial result, and estimating $p_1$ and $p_2$ after the trial has been completed should in any case supplement the result of the test in \Cref{eq:twoSidedParameterized} to provide a holistic interpretation of the test result. Another option would be to perform one of the directional tests
\begin{align}
    &H_0:\eta \leq 0 \hspace{1cm} \text{ versus } \hspace{1cm} H_1:\eta > 0\label{eq:directionalParameterized}\\
    &H_0:\eta = 0 \hspace{1cm} \text{ versus } \hspace{1cm} H_1:\eta > 0\label{eq:directionalParameterizedOneSided1}\\
    &H_0:\eta = 0 \hspace{1cm} \text{ versus } \hspace{1cm} H_1:\eta < 0\label{eq:directionalParameterizedOneSided2}
\end{align}
where $H_0$ in \Cref{eq:directionalParameterized} implies that the placebo or standard of care works at least as good as the novel treatment or drug, and $H_1$ can be interpreted as the novel drug having a larger efficacy than the placebo or standard of care. The test in \Cref{eq:directionalParameterizedOneSided1} assumes, on the other hand, that it is unrealistic a priori that $\eta < 0$. That is, $p_2<p_1$ is excluded a priori. This could be reasonable if the control arm includes a standard of care and the treatment arm the standard of care plus an addon treatment, which is known not to interfere with the standard of care. Therefore, the success probability in the treatment group should, theoretically, at least be equally large as in the control group. Likewise, \Cref{eq:directionalParameterizedOneSided2} can be useful in a setting where the binary endpoint measures failures. Then, excluding $\eta>0$ a priori as a possibility means that $p_2\leq p_1$ must hold. This is a reasonable assumption if again, the control group gets a standard of care and the treatment group the standard of care plus an addon treatment which is known not to interfere with the former. In this section, we focus solely on the two-sided test in \Cref{eq:twoSidedParameterized}, but we will derive Bayes factors, design and analysis priors and the required prior-predictives for power and sample size calculations as outlined in \Cref{fig:overview} for the tests in \Cref{eq:directionalParameterized}, \Cref{eq:directionalParameterizedOneSided1} and \Cref{eq:directionalParameterizedOneSided2} in \Cref{subsec:one-sided-hyp1}, \Cref{subsec:one-sided-hyp2} and \Cref{subsec:one-sided-hyp3}.

\subsubsection{Choice of the design and analysis prior}

The first step for the sample size calculation approach is shown as steps 1 and 2 in \Cref{fig:overview}. Thus, so called design priors and analysis priors must be chosen \citep{Grieve2022,KelterPawel2025,PawelHeld2025}. We denote the former by $P_{p_1}^d$ and $P_{p_2}^d$ and the latter by $P_{p_1}^{a}$ and $P_{p_2}^{a}$. The idea of the design prior is that we base the planning of the sample size and properties such as the Bayesian type-I-error rate and power -- compare \Cref{eq:bayesianPower} -- on this prior $P_{p_i}^d, i=1,2$. However, the planning stage might include subjective beliefs, so the analysis -- in this case a Bayes factor test -- should be based on another (more objective) prior distribution $P^{a}$ \citep{ohaganBayesianAssessmentSample2001}. When choosing $P_{p_i}^d = P_{p_i}^{a}$ for $i=1,2$, the design and analysis priors agree and we base both the planning and analysis of the study on the same distribution. In practice, different stakeholders will have different requirements on the analysis respectively design prior. For example, from a regulatory authorities' perspective, using a noninformative, 'objective' analysis prior will be almost mandatory. A highly informative, or 'subjective' analysis prior will be seen with suspicion. On the other hand, an investigator might be tempted to use a more informative analysis prior, maybe based on historical data, to make the trial require fewer patients. Regarding the design prior, during the planning stage of the trial a more informative design prior will cause less suspicion from a regulatory perspective than an informative analysis prior, as long as the long-term frequentist properties of the resulting design are calibrated in a reasonable sense. We will illustrate this aspect later in \Cref{subsec:examples}.

Now, we start with the design prior and must choose a prior under $H_0$ and under $H_1$, to use the prior-predictive density in a second step to calculate our desired quantities in \Cref{eq:bayesianDesign}. Thus, two priors $P_{p_1}^d$ and $P_{p_2}^d$ are chosen for the parameters $p_i \in [0,1]$, $i=1,2$. The $\mathrm{Beta}(a_0,b_0)$ distribution is a conjugate prior for the binomial likelihood, and when chosen as the prior, the posterior $P_{p_i \mid Y_i}$ is also Beta-distributed \citep{Held2014}:
\begin{align*}
	p_i \mid Y_i=y\sim \mathrm{Beta}(a_0+y,b_0+n-y)
\end{align*}
for $i=1,2$. As a consequence, a natural choice for the priors is the beta distribution. We follow \cite{Dickey1970a} and choose a Beta design prior under $H_0$ as follows:
\begin{align}\label{eq:designPriorH0twoSided}
    &p_1 =p_2 = p\mid H_0 \sim \mathrm{Beta}(a_0^d,b_0^d)
\end{align}
Thus, under $H_0:\eta = 0$, both probabilities are identical, $p_1=p_2$, and take some value $p\in [0,1]$, which has a beta design prior. Likewise, we pick independent Beta design priors under $H_1:\eta \neq 0$:
\begin{align}\label{eq:designPriorH1twoSided}
    &p_1 \mid H_1 \sim \mathrm{Beta}(a_1^d,b_1^d)\\
    &p_2 \mid H_1 \sim \mathrm{Beta}(a_2^d,b_2^d)
\end{align}
For the analysis priors $P_{p_1}^a$, $P_{p_2}^a$ under $H_1$, we also choose independent Beta priors, with possibly different values $a_i^a$ and $b_i^a$ for $i=1,2$, where the superscript signals that the hyperparameters belong to our analysis instead of design prior:
\begin{align}
    &p_1 \mid H_1 \sim \mathrm{Beta}(a_1^a,b_1^a)\label{eq:analysisPriorsTwoSidedH0}\\
    &p_2 \mid H_1 \sim \mathrm{Beta}(a_1^a,b_1^a)\label{eq:analysisPriorsTwoSidedH1}
\end{align}
Lastly, for the analysis prior $P_{p}^a$ under $H_0:\eta=0$, we choose a Dirac prior with all probability on $\eta=p_2-p_1=0$ conditionally on a uniform prior on the grand mean $\zeta$, that is
$$p_1=p_2=p|H_0 \sim 1_{\{\eta=0\}}| \zeta \sim U(0,1)$$
for the analysis with the Bayes factor. Thus, we remain vague and flexible with independent beta design priors during the planning stage of the trial, and pick likewise flexible priors under $H_0$ and $H_1$ to carry out a two-sided Bayes factor test in the beta-binomial model for the analysis stage \citep{Dickey1970a,Jamil2017}.

\subsubsection{Derivation of the prior-predictive distribution}
The third step shown in \Cref{fig:overview} consists of deriving the prior-predictive probability mass function $f(y \mid H_0)$ and $f(y \mid H_1)$ under the null and alternative hypothesis. These probability mass functions will be used together with the Bayes factor $\mathrm{BF}_{01}(y)$ later, to compute critical values $y_{crit}$, for which we can state that $\mathrm{BF}_{01}$ passes a given threshold such as $k$. Based on the predictive distributions we can then compute the desired probabilities $P(\mathrm{BF}_{01}(y)<k \mid H_0)$ and $P(\mathrm{BF}_{01}(y)<k \mid H_1)$, compare \Cref{eq:bayesianDesign}. Achieving a specified power will then proceed by selecting the smallest sample size $n\in \mathbb{N}$ for which the predictive probability satisfies
$$P(y>y_{crit} \mid H_1)\geq 1-\beta$$
for a specified $\beta$ such as $\beta:=0.2$, where $P(\cdot  \mid H_1)$ denotes the predictive distribution under $H_1$.

We base the predictive probability mass function on two observed samples $Y_1$ (binary data of the control group) and $Y_2$ (binary data of the treatment group) as well as the independent beta design priors under $H_0$ and $H_1$ to obtain a prior-predictive distribution. This leads to the following predictive probability mass function of the data $D=(Y_1,Y_2)$ based on fixed $n_1,n_2$ and independent beta priors $p_1 \sim \mathrm{Beta}(a_1^d, b_1^d)$, for details see the Appendix:
\begin{align}\label{eq:predDensityH1}
        f(d|&H_1)={n_1 \choose y_1}\frac{\mathrm{B(y_1+a_1^d,n_1-y_1+b_1^d)}}{\mathrm{B}(a_1^d,b_1^d)} \cdot  {n_2 \choose y_2} \frac{\mathrm{B(y_2+a_2^d,n_2-y_2+b_2^d)}}{\mathrm{B}(a_2^d,b_2^d)}
\end{align}
Using the properties of the Beta function $\mathrm{B}(z_1,z_2):=\int_0^1 t^{z-1}(1-t)^{z_2-1}$, namely $\mathrm{B}(z_1,z_2)=\frac{\Gamma(z_1)\Gamma(z_2)}{\Gamma(z_1+z_2)}$ where $\Gamma(n):=(n-1)!$ denotes the Gamma function, the above can also be written as
\begin{align}\label{eq:predDensityGammaRepresentation}
    &=\frac{\Gamma(n_1+1)\Gamma(y_1+a_1^d)\Gamma(n_1-y_1+b_1^d)}{\Gamma(n_1+a_1^d+b_1^d)\Gamma(y_1+1)\Gamma(n_1-y_1+1)}\frac{\Gamma(a_1^d+b_1^d)}{\Gamma(a_1^d)\Gamma(b_1^d)}\nonumber\\
    &\cdot \frac{\Gamma(n_2+1)\Gamma(y_2+a_2^d)\Gamma(n_2-y_2+b_2^d)}{\Gamma(n_2+a_2^d+b_2^d)\Gamma(y_2+1)\Gamma(n_2-y_2+1)}\frac{\Gamma(a_2^d+b_2^d)}{\Gamma(a_2^d)\Gamma(b_2^d)}
\end{align}
Depending on which design priors we use, we further replace $a_1^d$ and $b_1^d$ as well as $a_2^d$ and $b_2^d$ by the selected values in the design prior under $H_1$ in \Cref{eq:designPriorH1twoSided}. We stress that the prior-predictive distribution in \Cref{eq:predDensityH1} makes use of the design priors hyperparameters chosen in advance. The prior-predictive density under $H_0$ can be derived as
\begin{align}\label{eq:predDensityH0}
    f(d|H_0)=\frac{{n_1 \choose y_1}{n_2 \choose y_2}}{\mathrm{B}(a_0^d,b_0^d)} \mathrm{B}(a_0^d+y_1+y_2,b_0^d+n_1+n_2-y_1-y_2)
\end{align}
for details see the Appendix.

\subsubsection{Derivation of the Bayes factor}
The next and fourth step in \Cref{fig:overview} is the derivation of the Bayes factor. We use the independent beta analysis priors discussed in the previous subsection. Still, we stress that it is well possible to choose different hyperparameters $a_i^a$ and $b_i^a$ for $i=1,2$ in the analysis than in the design priors $P_{p_i}^d$ (where the hyperparameters were $a_i^d$ and $b_i^d$ for $i=1,2$). It is, of course, also possible to choose design and analysis priors identically by using the same hyperparameters. Using the Savage-Dickey-density-ratio, \cite[p.~219]{Dickey1970a} derived the Bayes factor for this setting analytically as:
\begin{align}\label{eq:bf_twoSided}
    \mathrm{BF}_{01}(y_1,y_2):&=\frac{\mathrm{B}(a_1^a+y_1+a_2^a+y_2-1,b_1^a+n_1+b_2^a+n_2-y_1-y_2-1)\mathrm{B}(a_1^a,b_1^a)\mathrm{B}(a_2^a,b_2^a)}{\mathrm{B}(a_1^a+y_1,b_1^a+n_1-y_1)\mathrm{B}(a_2^a+y_2,b_2^a+n_2-y_2)\mathrm{B}(a_1^a+a_2^a-1,b_1^a+b_2^a-1)}
\end{align}
$\mathrm{BF}_{01}(y_1,y_2)$ is a function of the observed successes $y_1$ in the control group and $y_2$ in the treatment group. Once the sample sizes $n_1$ and $n_2$ in both groups are fixed and the analysis prior hyperparameters $a_i^a$ and $b_i^a$ are fixed for $i=1,2$, the Bayes factor testing \Cref{eq:twoSidedParameterized} can be computed. It is worth noting that the same Bayes factor is obtained when testing for independency of rows and columns in a $2\times 2$ contingency table, once either the rows or columns are fixed \citep{Gunel1974,Jamil2017}. \Cref{tab:2x2_table} shows the structure of such a $2\times 2$ contingency table, and in the context of a phase-II-b trial with treatment and control group, the sample size of the groups, and thereby the sample size of the rows, is fixed in advance as $n_1$ and $n_2$ after the sample size calculation for the trial has been carried out. 

\begin{table}[h!]
    \centering
    \begin{tabular}{c|cc|c}
         & success & failure & $\sum$ \\
         \hline
         treatment & $y_2$ & $n_2-y_2$ & $n_2$\\
         control & $y_1$ & $n_1-y_1$ & $n_1$\\
         \hline
         $\sum$ & unknown & unknown & $\sum=n_1+n_2$\\
    \end{tabular}
    \caption{Structure of the $2\times 2$ contingency table for which the Bayes factor in \Cref{eq:bf_twoSided} can test for row-column independence when either the rows or columns are fixed.}
    \label{tab:2x2_table}
\end{table}
\Cref{tab:2x2_table} shows that our approach can be used also in settings where the goal might not be the test of $H_0:\eta=0$ against $H_1:\eta \neq 0$, but the test of stochastic independence between rows and columns, e.g. in epidemiological studies without randomization into treatment and control group.

\subsubsection{Numerical root-finding}\label{subsubsec:rootfinding}
Based on the Bayes factor, the fifth step in \Cref{fig:overview} now consists of numerical root-finding. In brief terms, the idea is to find a solution to the equation
\begin{align}\label{eq:rootfinding}
    \mathrm{BF}_{01}(y_1,y_2)=k
\end{align}
by numerical means for fixed sample sizes $n_1,n_2$. However, using \Cref{eq:bf_twoSided} and numerically finding the root of $\mathrm{BF}_{01}(y_1,y_2)-k=0$ via Newton's method as, for example, implemented in the \texttt{uniroot} function in the statistical programming language R \citep{RProgrammingLanguage} is not straightforward, because there is more than a single solution $(y_1,y_2)$. 

\begin{figure}[h!]
    \centering
    \includegraphics[width=0.5\linewidth]{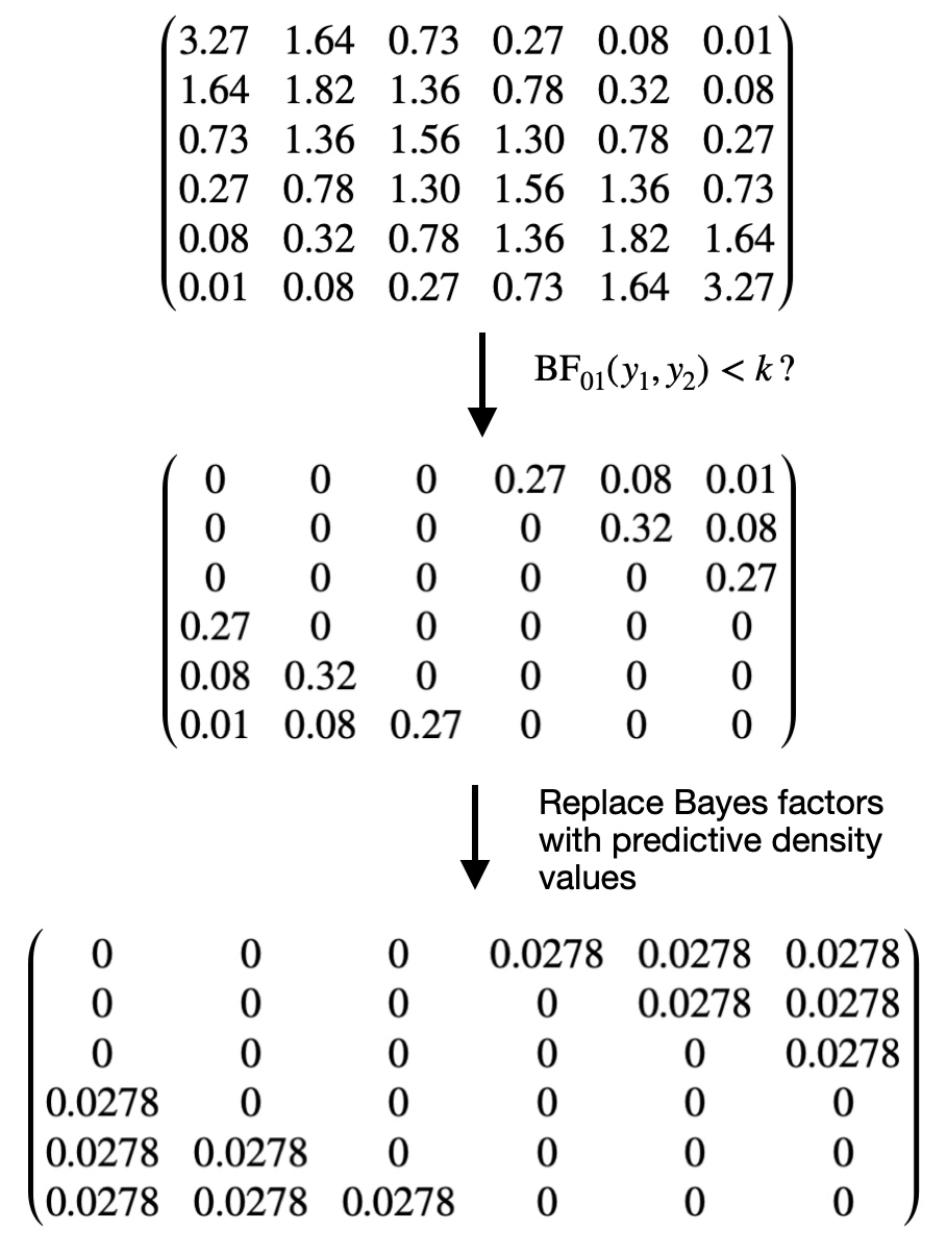}
    \caption{Illustration of the computation of critical values for the Bayes factor power calculation, based on $n_1=n_2=5$ and flat analysis priors with $a_i^a=b_i^a=1$ for $i=1,2$.}
    \label{fig:matrixWorkflow}
\end{figure}

For illustration purposes, suppose $n_1=n_2=5$. Then, $y_i=0,...,5$ for $i=1,2$, as in both groups zero to five successes are observable. The top matrix in \Cref{fig:matrixWorkflow} visualizes the possible results as a $6\times 6$ matrix, where the rows denote the number of successes $y_1$ in the control group, and the columns the number of successes $y_2$ in the treatment group. Thus, the top-left entry $3.27$ is the Bayes factor $\mathrm{BF}_{01}(y_1=0,y_2=0)$, the entry right of that $\mathrm{BF}_{01}(y_1=0,y_2=1)=1.64$ and so on. The top matrix in \Cref{fig:matrixWorkflow} shows the resulting Bayes factors $\mathrm{BF}_{01}(y_1,y_2)$ for all values $y_i=0,...,5$ for $i=1,2$, where flat analysis priors with $a_i^a=b_i^a=1$ have been chosen in both groups for $i=1,2$. Now, two phenomena can be observed from the top matrix in \Cref{fig:matrixWorkflow}: First, the matrix is symmetric, which is due to the flat priors and the two-sided test of $H_0:\eta=0$ versus $H_1:\eta \neq 0$. Second, the largest Bayes factors are always obtained in the main diagonal, as there $y_1=y_2$ holds and the number of successes is equal in both groups, providing evidence for $H_0:\eta = 0$. For a fixed row or column, the farther away from the main diagonal entry one moves horizontally in the row or vertically in the column, the smaller the Bayes factor becomes. This is to be expected, as the difference between $y_1$ and $y_2$ becomes more pronounced in this case. These two phenomena illustrate that small Bayes factors which signal evidence against $H_0:\eta=0$ can be found primarily in the top-right and bottom-left margins of the matrix under flat design priors.

\subsubsection{Computation of critical value(s)}
In the second matrix in the middle of \Cref{fig:matrixWorkflow}, all entries for which the Bayes factor does not pass the requires evidence threshold $k=1/3$ have been set to zero. Thus, only entries remain for which the tupel $(y_1,y_2)$ yields a result which would lead to the rejection of $H_0:\eta = 0$ by means of $\mathrm{BF}_{01}(y_1,y_2)<1/3$ (which implies $\mathrm{BF}_{10}(y_1,y_2)>3$, indicating at least moderate evidence in favour of $H_1:\eta \neq 0$ according to \cite{jeffreys1961}). The non-zero entries therefore are the ones for which the critical values $(y_1,y_2)$ lead to sufficient evidence $1/k$ in favour of $H_1:\eta \neq 0$ based on the Bayes factor. We denote the set of these tupels as $\mathcal{Y}$ henceforth. In the example of \Cref{fig:matrixWorkflow}, we have $\mathcal{Y}=\{(y_1=0,y_2=3), (y_1=0,y_2=4), (y_1=0,y_2=5), (y_1=1,y_2=4), (y_1=1,y_2=5), (y_1=2,y_2=5), (y_1=3,y_2=0), (y_1=4,y_2=0), (y_1=5,y_2=0), (y_1=4,y_2=1), (y_1=5,y_2=1), (y_1=5,y_2=2)\}$.

\subsubsection{Computation of Bayesian type-I-error rate and power}
Based on the critical values found in the last step, the seventh step in \Cref{fig:overview} consists of computing the relevant quantities
$P(\mathrm{BF}_{01}(y)<k \mid H_0)$ and $P(\mathrm{BF}_{01}(y)<k \mid H_1)$
where we have
$$P(\mathrm{BF}_{01}(y)<k \mid H_0)=P(\mathcal{Y} \mid H_0)$$
and
$$P(\mathrm{BF}_{01}(y)<k \mid H_1)=P(\mathcal{Y} \mid H_1)$$
and the former is equal to the Bayesian type-I-error rate, while the latter is the Bayesian analogue to frequentist power, compare \Cref{eq:bayesianDesign}.

Therefore, compare the bottom matrix in \Cref{fig:matrixWorkflow}. The Bayes factor values $\mathrm{BF}_{01}(y_1,y_2)$ have been replaced by the values of the predictive density $f(y_1,y_2 \mid n_1, n_2, a_1^d, b_1^d, a_2^d, b_2^d)$ derived in \Cref{eq:predDensityH1}. In the last step, these are summed to obtain the predictive probability $P(\mathrm{BF}_{01}(y)<k \mid H_1)$. For example, in the setting of \Cref{fig:matrixWorkflow} we obtain the Bayesian power
\begin{align}
    P(\mathrm{BF}_{01}(y)<k \mid H_1)=P(\mathcal{Y} \mid H_1)&=\sum_{(y_1,y_2)\in \mathcal{Y}}f(y_1,y_2 \mid n_1, n_2, a_1^d, b_1^d, a_2^d, b_2^d)\nonumber\\
    &= 12\cdot 0.0278 \approx 0.33\nonumber
\end{align}
Likewise, we can compute the Bayesian type-I-error rate by using $f(y_1,y_2|H_0)$ instead of $f(y_1,y_2|H_1)$ in the above.
\subsubsection{Sample size calculation for the Bayes factor}\label{subsubsec:samplesizecalc}
The ultimate goal now is to obtain the sample size $n$ for which we can state that $P(\mathrm{BF}_{01}(y)<k \mid H_1)=P(\mathcal{Y} \mid H_1)$ exceeds a given threshold, such as $1-\beta$ for $\beta:=0.2$, so we have at least $80\%$ Bayesian power to find at least evidence $1/k$ for $H_1$ (or likewise, evidence $k$ against $H_0$). It might happen that for a fixed $n_1$ and $n_2$ the set $\mathcal{Y}$ is not large enough so that the threshold $\mathrm{BF}_{01}(y)< k$ is fulfilled for enough entries and the resulting $\sum_{(y_1,y_2)\in \mathcal{Y}}f(y_1,y_2 \mid n_1, n_2, a_1^d, b_1^d, a_2^d, b_2^d)$ becomes larger than the desired threshold $1-\beta$ (e.g. 80\%). In the most extreme case, the set $\mathcal{Y}$ might even become empty, as no combination of $y_1$ and $y_2$ values produces a Bayes factor $\mathrm{BF}_{01}(y)< k$ (in this case, in \Cref{fig:matrixWorkflow} the middle matrix would consist only of zeros). However, in such cases, increasing $n_1$ and $n_2$ will eventually lead to the situation where a small enough Bayes factor can be found. This is due to the fact that for large enough sample sizes $n_1$ and $n_2$, there must be a number of successes $y_1$ and $y_2$ for which $\mathrm{BF}_{01}(y_1,y_2)<k$ holds due to the asymptotic properties of the Bayes factor and the consistency of the posterior distribution \citep{Kleijn2022}.

Still, \Cref{fig:overview} illustrates that under noninformative, flat design priors, the probability mass which contributes to the power is located primarily in the top-right and bottom-left margins of the corresponding matrix which represents the situation. For large $n_1$ and $n_2$, the predictive probabilities in these margins will shrink more and more, and thus it might be possible, that under an entirely noninformative design prior choice, the power will increase only very slowly. Thus, it might be reasonable to separate the hypotheses $H_0:\eta=0$ and $H_1:\eta \neq 0$ by using more informative design priors. We will illustrate this phenomenon later in \Cref{subsec:examples}.

Taking stock, we have chosen design and analysis priors, derived the Bayes factor and developed a simple method to replace standard numerical root-finding with a matrix-search-algorithm. Computation of the critical values is based on this approach and the computation of Bayesian type-I-error rate and power then makes use of the design and analysis priors and Bayes factor. Together, this allows to repeat these steps for increasing sample sizes $(n_1,n_2)$ and investigate when the Bayesian power, type-I-error rate or probability of compelling evidence reach a desired level. We close this subsection by noting that no asymptotic theory or Monte Carlo simulations are required to do so, solely standard numerical methods like numerical integration.

\subsection{One-sided hypothesis test of $H_0:\eta=0$ versus $H_+:\eta>0$}\label{subsec:one-sided-hyp1}
The last section illustrated all eight steps of our power and sample size calculation approach for Bayes factors in the two-arm binomial setting shown in \Cref{fig:overview}. In the last section, we focussed on the two-sided test of $H_0:\eta=0$ versus $H_1:\eta \neq 0$. This section provides the analogue derivations for the one-sided test of $H_0:\eta=0$ versus $H_+:\eta>0$.

\subsubsection{Binomial model and hypotheses}

We observe
\[
Y_1 \sim \mathrm{Bin}(n_1,p_1),
\qquad
Y_2 \sim \mathrm{Bin}(n_2,p_2),
\]
and wish to test
\[
H_0 : \eta = 0 \quad\text{vs.}\quad H_+ : \eta > 0,
\qquad
\eta = p_2 - p_1.
\]
Under \(H_0\) we have \(p_1 = p_2 =: p\).
Under \(H_+\) we impose the order constraint \(p_2 > p_1\).

\subsubsection{Design and analysis priors under \(H_0\) and \(H_+\)}

The Beta density is given as
\[
\pi_{\text{Beta}}(p \mid a,b)
= \frac{1}{B(a,b)}\,p^{a-1}(1-p)^{b-1},
\qquad 0 < p < 1,
\]
where the Beta function is
\[
B(a,b)
= \int_0^1 t^{a-1}(1-t)^{b-1}\,dt
= \frac{\Gamma(a)\Gamma(b)}{\Gamma(a+b)}.
\]
The (unregularized) incomplete Beta function and its regularized version are
\[
B_x(a,b)
= \int_0^x t^{a-1}(1-t)^{b-1}\,dt,
\qquad
I_x(a,b)
= \frac{B_x(a,b)}{B(a,b)}.
\]
Under \(H_0:p_1=p_2=:p\) we take
\[
p \sim \mathrm{Beta}(a_0^d,b_0^d),
\]
with prior density
\[
\pi_0(p)
= \frac{1}{B(a_0^d,b_0^d)}\,p^{a_0^d-1}(1-p)^{b_0^d-1},
\qquad 0<p<1.
\]
where the superscript $d$ indicates that this is the prior used in the predictive density under $H_0$. Likewise, the analysis prior used when computing a Bayes factor is also Beta distributed,
\[
p \sim \mathrm{Beta}(a_0^a,b_0^a),
\]
where the superscripts $a$ indicate the analysis prior. Under \(H_+\) we start from independent Beta design priors
\[
p_1 \sim \mathrm{Beta}(a_1^d,b_1^d),
\qquad
p_2 \sim \mathrm{Beta}(a_2^d,b_2^d),
\]
with joint (untruncated) density
\[
\pi_{\text{untr}}(p_1,p_2)
= \frac{1}{B(a_1^d,b_1^d)B(a_2^d,b_2^d)}\,
p_1^{a_1^d-1}(1-p_1)^{b_1^d-1}\,
p_2^{a_2^d-1}(1-p_2)^{b_2^d-1},
\qquad 0<p_1,p_2<1.
\]
We then truncate to the order-restricted region
\[
A = \{(p_1,p_2)\colon 0 < p_1 < p_2 < 1\}.
\]
The normalizing constant for the truncated prior is
\[
C
= \iint_{A} \pi_{\text{untr}}(p_1,p_2)\,dp_1\,dp_2
= P(p_2>p_1)
\]
under the independent Beta\((a_1^d,b_1^d)\) and Beta\((a_2^d,b_2^d)\) priors. A convenient expression in terms of \(I_x(a,b)\) is
\begin{align}\label{eq:C}
C
= \int_0^1
\frac{p_2^{a_2^d-1}(1-p_2)^{b_2^d-1}}{B(a_2^d,b_2^d)}\,
I_{p_2}(a_1^d,b_1^d)\,dp_2.
\end{align}
Hence the truncated prior density under \(H_+\) is
\[
\pi_+(p_1,p_2)
= \frac{\pi_{\text{untr}}(p_1,p_2)\,\mathbf{1}\{0<p_1<p_2<1\}}{C}.
\]

\begin{figure}[h!]
    \centering
    \includegraphics[width=1\linewidth]{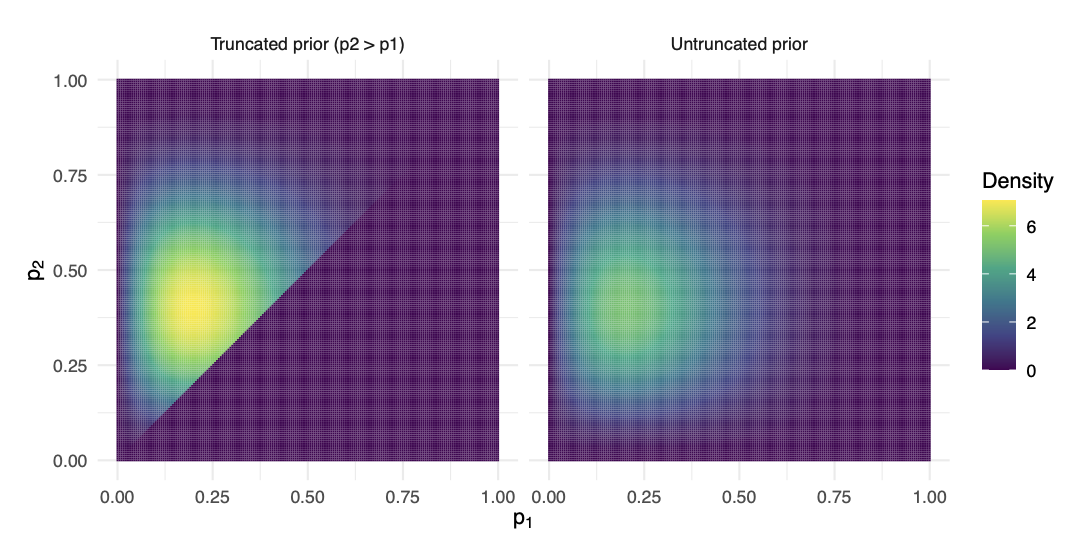}
    \caption{Untruncated Beta$(a_1^d,a_2^d)$ design (or analysis) prior (right) and truncated Beta$(a_1^d,a_2^d)$ prior (left) for $a_1^d=2$, $b_1^d=5$, $a_2^d=3$ and $b_2^d=4$, where the truncation is to the set $\iint_{A} \pi_{\text{untr}}(p_1,p_2)\,dp_1\,dp_2
= P(p_2>p_1)$.}
    \label{fig:priors}
\end{figure}
\Cref{fig:priors} shows an example of the untruncated and truncated priors under $H_+$. The right panel in \Cref{fig:priors} shows that the probability mass under the untruncated prior distributes on the whole set $[0,1]\times[0,1]$, while the left panel only has positive probability mass on the set $A$ anymore, which is in line with the restriction $p_2>p_1$ imposed under $H_+:\eta >0$.

\subsubsection{Finite-sum form for \(C\) (integer shapes)}

When \(a_1^d\) and \(a_2^d\) are positive integers, we can write a finite-sum expression
for \(C\) and, therefore, for $p(y_1,y_2|H_1)$, too. For details see the Appendix.

\subsubsection{Predictive density under \(H_0\)}

Under \(H_0\), \(p_1=p_2=p\) and, conditional on \(p\),
\[
Y_1 \mid p \sim \mathrm{Bin}(n_1,p),
\qquad
Y_2 \mid p \sim \mathrm{Bin}(n_2,p),
\]
independently. The probability mass function given \(p\) is
\[
f(p \mid y_1,y_2)
= \binom{n_1}{y_1}\binom{n_2}{y_2}
p^{y_1+y_2}(1-p)^{(n_1-y_1)+(n_2-y_2)}.
\]
With the prior \(p \sim \mathrm{Beta}(a_0^d,b_0^d)\),
\[
\pi_0(p)
= \frac{1}{B(a_0^d,b_0^d)}\,p^{a_0^d-1}(1-p)^{b_0^d-1},
\]
the predictive density (marginal likelihood) under \(H_0\) is
\begin{align}
p(y_1,y_2 \mid H_0)
&= \int_0^1
f(p \mid y_1,y_2)\,\pi_0(p)\,dp \nonumber\\
&= \binom{n_1}{y_1}\binom{n_2}{y_2}
\int_0^1
\frac{1}{B(a_0^d,b_0^d)}\,
p^{y_1+y_2+a_0^d-1}(1-p)^{n_1+n_2-y_1-y_2+b_0^d-1}\,dp \nonumber\\
&= \binom{n_1}{y_1}\binom{n_2}{y_2}
\frac{B\bigl(y_1+y_2+a_0^d,\;n_1+n_2-y_1-y_2+b_0^d\bigr)}{B(a_0^d,b_0^d)}.\label{eq:predDensH0}
\end{align}

\subsubsection{Predictive density under \(H_+\)}

Given \((p_1,p_2)\), the probability mass function factorizes as
\[
f(p_1,p_2 \mid y_1,y_2)
= \binom{n_1}{y_1}\binom{n_2}{y_2}
p_1^{y_1}(1-p_1)^{n_1-y_1}
p_2^{y_2}(1-p_2)^{n_2-y_2}.
\]
The predictive density under \(H_+\) can be derived as
\begin{align}\label{eq:predDensH+}
    p(y_1,y_2 \mid H_+)
= \binom{n_1}{y_1}\binom{n_2}{y_2}
\frac{I(y_1,y_2)}{B(a_1^d,b_1^d)B(a_2^d,b_2^d)\,C}.
\end{align}
with the updated shape parameters
\begin{align}\label{eq:updatedShapeParameters}
A_1 = y_1 + a_1^d,\quad B_1 = n_1 - y_1 + b_1^d,\qquad
A_2 = y_2 + a_2^d,\quad B_2 = n_2 - y_2 + b_2^d.
\end{align}
    where
\[
J
= \int_0^1
p_1^{A_1-1}(1-p_1)^{B_1-1}I_{p_1}(A_2,B_2)\,dp_1 \hspace{1cm}\text{ and } \hspace{1cm}I(y_1,y_2)
= B(A_2,B_2)\bigl[B(A_1,B_1) - J\bigr]
\]
for details see the Appendix.

\subsubsection{Finite-sum form for \(I(y_1,y_2)\) (integer shapes)}

When \(A_2\) is a positive integer, a finite-sum form for $I(y_1,y_2)$ is available again, for details see the Appendix.

\subsubsection{Bayes factor for $H_0:\eta=0$ versus $H_+:\eta>0$}

The Bayes factor for \(H_+\) versus \(H_0\) is
\[
\mathrm{BF}_{+0}(y_1,y_2)
= \frac{p(y_1,y_2 \mid H_+)}{p(y_1,y_2 \mid H_0)},
\]
with \(p(y_1,y_2 \mid H_0)\) and \(p(y_1,y_2 \mid H_+)\) as derived above and can be computed by calculating \Cref{eq:predDensH+} and dividing it by \Cref{eq:predDensH0}:
\begin{align}\label{eq:BF+0}
   \mathrm{BF}_{+0}(y_1,y_2)
&= \frac{\binom{n_1}{y_1}\binom{n_2}{y_2}
\frac{I(y_1,y_2)}{B(a_1^a,b_1^a)B(a_2^a,b_2^a)\,C}}{\binom{n_1}{y_1}\binom{n_2}{y_2}
\frac{B(y_1+y_2+a_0^a,\;n_1+n_2-y_1-y_2+b_0^a)}{B(a_0^a,b_0^a)}}\nonumber\\
&=\frac{B(a_0^a,b_0^a)I(y_1,y_2)}{B(y_1+y_2+a_0^a,\;n_1+n_2-y_1-y_2+b_0^a)B(a_1^a,b_1^a)B(a_2^a,b_2^a)C}\nonumber\\
&=\frac{B(a_0^a,b_0^a)B(A_2,B_2)\bigl[B(A_1,B_1) - J\bigr]}{B(y_1+y_2+a_0^a,\;n_1+n_2-y_1-y_2+b_0^a)B(a_1^a,b_1^a)B(a_2^a,b_2^a)C}\nonumber\\
&=\frac{B(a_0^a,b_0^a)B(A_2,B_2)\bigl[B(A_1,B_1) - \int_0^1
p_1^{A_1-1}(1-p_1)^{B_1-1}I_{p_1}(A_2,B_2)\,dp_1\bigr]}{B(y_1+y_2+a_0^a,\;n_1+n_2-y_1-y_2+b_0^a)B(a_1^a,b_1^a)B(a_2^a,b_2^a)C}
\end{align}
where either \Cref{eq:C} or, in case integer values are used for the shape parameters of the truncated beta analysis priors, \Cref{eq:finiteSumFormC} in the Appendix can be substituted for $C$ in the denominator of \Cref{eq:BF+0}. Note that we have replaced the hyperparameters of the design priors, $a_i^d$, $b_i^d$ in the predictive densities with the ones of a possibly different analysis prior used to calculate the Bayes factor, $a_i^a$, $b_i^a$ in \Cref{eq:BF+0} above. Note also that the observed data $y_1,y_2$ as well as the analysis prior hyperparameters $a_1^a,b_1^a,a_2^a,b_2^a$ influence the Bayes factor in form of the updated shape parameters as given in \Cref{eq:updatedShapeParameters}. Computationally, the Bayes factor $\mathrm{BF}_{+0}$ can be computed by numerical integration and with Beta functions. When the shape parameters are integers, no numerical computation for the calculation of $C$ is necessary anymore, further simplifying the computational effort.

\subsection{One-sided hypothesis test of $H_0:\eta=0$ versus $H_-:\eta<0$}\label{subsec:one-sided-hyp2}

The one-sided hypothesis test of $H_0:\eta=0$ versus $H_-:\eta<0$ is very similar to the test of $H_0:\eta=0$ versus $H_-:\eta>0$ treated above. Therefore, we relegate the required derivations fully into the Appendix, and summarize only the main differences. First, the truncation area $A$ changes accordingly as the order restriction imposed now is $p_1 < p_2$, which influences the predictive area under $H_-:\eta <0$. The predictive density under $H_0$ remains unchanged, and the Bayes factor again can be obtained via standard numerical integration.

\subsection{One-sided hypothesis test of $H_0:\eta \leq 0$ versus $H_1:\eta >0$}\label{subsec:one-sided-hyp3}

The one-sided hypothesis test of $H_0:\eta \leq 0$ versus $H_1:\eta >0$ is similar to the test of $H_0:\eta=0$ versus $H_-:\eta>0$ treated above, too. Therefore, we relegate the required derivations fully into the Appendix. Again, the truncation areas of the design and analysis priors are influenced according to the imposed order constraints under $H_+$ and $H_-$. That is, under $H_+$ we restrict the prior to $p_2 > p_1$, while under $H_-$, we restrict the prior to $p_1 \geq p_2$.

\subsection{Real-World Numerical Examples from Phase II Trials}\label{subsec:examples}

The last section provided a walk-through and all relevant details to the power and sample size calculation method we propose for Bayes factors in two-arm clinical trials with binary endpoints. As the last sections clarify, we do not make use of simulating trial data under $H_0$ or $H_1$. Also, the proposed method does not rely on asymptotic theory, so the provided sample sizes and power values are precise. In this section, we illustrate the application of our method for two-arm phase II trials with binary endpoints. Therefore, we use data from two published randomized controlled trials where exact counts of successes and failures are reported per arm. We provide a reanalysis of both trials and interpret the output of our approach, showing how beneficial our method is when conducted in addition to a standard Bayesian analysis. Also, we clarify how planning a trial from scratch to achieve a desired power and type-I-error rate is straightforward with our method.\footnote{All analyses and plots can be recreated using the R package \texttt{binbf2arm}, and running the code in our Quarto replication script available at \url{https://osf.io/zsrfh/overview?view_only=351aecc896d0468991569c99ae58beaf}. The R package is available on CRAN under \url{https://cran.r-project.org/web/packages/bfbin2arm/index.html}.}

\subsubsection{Riociguat phase II trial}
The riociguat trial is a phase IIb study in early diffuse cutaneous systemic sclerosis, which evaluated efficacy and safety of riociguat over 52 weeks \citep{khannaRiociguatPatientsEarly2020}. The primary endpoint was modified Rodnan Skin Score (mRSS) progression, where a failure is defined as an increase $>5$ units and $\geq 25\%$ from baseline at week 52. A success is defined as no progression.
\begin{table}[h]
\centering
\caption{Modified Rodnan Skin Score (mRSS) progression ($>5$ units and $\geq$25\%) at week 52 in RISE-SSc phase IIb trial}
\label{tab:riociguat-mrss}
\begin{tabular}{|l|c|c|}
\hline
 & Progression & No progression \\
\hline
Riociguat (n=59) & 11 & 48 \\
\hline
Placebo (n=60) & 22 & 38 \\
\hline
\end{tabular}
\end{table}
These counts -- see \Cref{tab:riociguat-mrss} -- (18.6\% vs 36.7\%) indicate a treatment benefit (OR 0.40, nominal $p=0.024$) \citep{khannaRiociguatPatientsEarly2020}.

The trial tested superiority of riociguat, resembling $H_0: p_1 = p_2$ vs. $H_+: p_1 < p_2$. For the endpoint skin progression, a success is defined as no progression, and thus evidence for $H_+$ means treatment with riociguat may lead to a higher proportion of patients with no progression (or equivalently, a lower proportion of patients with skin progression). Thus, ideally trial data provide evidence for $H_+$.

We use the theory derived in \Cref{subsec:one-sided-hyp1} and use moderate thresholds both for the Bayesian power and the probability of compelling evidence under $H_0$, that is, $k=1/3$ and $k_f=3$ in \Cref{eq:bayesianDesign} and \Cref{eq:pce_H0}.

\begin{figure}[h!]
    \centering
    \includegraphics[width=0.9\linewidth]{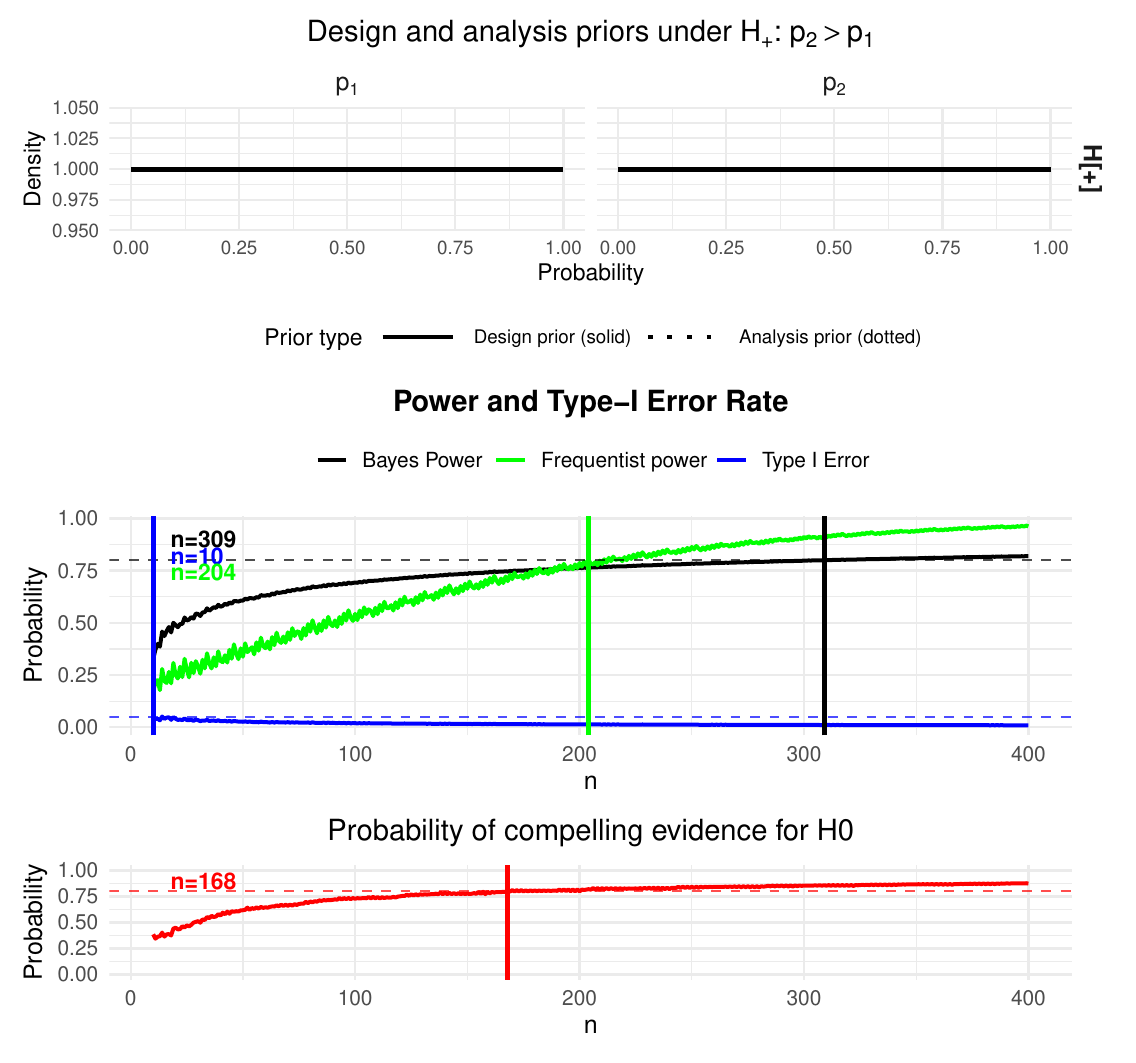}
    \caption{Bayesian power and sample size calculations for the riociguat trial, where $H_0:p_1=p_2$ versus $H_1:p_2>p_1$ is tested. Flat design and analysis priors are used with moderate evidence thresholds $k=1/3$, $k_f=3$. The calibration shows the results for 80\% power, 5\% type-I-error rate and 80\% probability of compelling evidence under $H_0$. Frequentist power was obtained under $p_1=0.4$ and $p_2=0.6$.}
    \label{fig:riociguat_flat}
\end{figure}

We use independent uniform Beta$(1,1)$ priors for each arm in the analysis (and a uniform Beta$(1,1)$ prior for the common-$p$ null model). Under these priors the Bayes factor $\mathrm{BF}_{+0}$ in favor of the null results in $\mathrm{BF}_{+0}=4.32$, indicating moderate evidence in favour of $H_+:p_2>p_1$ according to the scale of \cite{Jeffreys1939}. Therefore, one can would be tempted to conclude that progression occurs less frequent under riociguat treatment compared to placebo. Although this conclusion is allowed, there are no power or error guarantees associated with it, which might be problematic from a regulatory perspective. Such a perspective might require type-I-error control or a minimum prespecified power, compare \cite{FDABayes2010,ema2022reflectionpaper} and \cite{ionanBayesianMethodsHuman2023}. Thus, three questions emerge:
\begin{enumerate}
    \item{How many patients per arm need to be recruited to guarantee a minimum power of, say, $80\%$, and a type-I-error rate of, say, $5\%$? Did the trial recruit sufficient patients to assert long-term error guarantees for power and type-I-errors?}
    \item{How large is the Bayesian power and type-I-error rate based on the $119$ patients recruited for the trial? That is, based on $59$ patients in the treatment arm with $48$ successes and $60$ patients in the control arm with $38$ successes, we might be interested in the resulting power based on these sample sizes.}
    \item{How large is a \textit{fully frequentist} type-I-error rate (or power) of a design calibrated with our approach, in contrast to the Bayesian analogues of power and type-I-error rate in \Cref{eq:bayesianDesign}? From a regulatory perspective, a fully frequentist calibration might be required. Providing Bayesian analogues of power and type-I-error rate may not be sufficient.}
\end{enumerate}
\Cref{fig:riociguat_flat} shows the results of our power and sample size calibration with regard to the first question. The top plot shows the design and analysis priors used for $p_1$ and $p_2$, both of which are uniform, and thus, noninformative (the solid and dashed lines overlap in the top plots, which is why the design and analysis priors are visually not separable in this case). The middle plot provides the Bayesian power and type-I-error rate as well as the frequentist power (more on that below) as a function of the sample size.\footnote{Importantly, due to the fluctuations in the beta-binomial model, which are due to the discreteness of the data -- see also \cite{KelterPawel2025} for details -- the calibration ensures that the power does not drop below the desired threshold for the next ten sample sizes (likewise for the type-I-error rate, which is ensured not to become larger than the specified threshold for the next ten sample sizes).} Thereby, the middle plot shows the key result of our power and sample size calculation method outlined in \Cref{sec:methodology}. The bottom plot then shows the probability of compelling evidence -- see \Cref{eq:pce_H0} -- as a function of the sample size.

Now, regarding the first question above, the middle plot in \Cref{fig:riociguat_flat} shows that a power of $0.80$ is achieved at $n=309$, a Bayesian type-I error $\leq 0.05$ is achieved already at $n=10$, and a probability of compelling evidence for $H_0$, $P(\mathrm{BF}_{+0}>k_f|H_0) \geq 0.80$ is reached at $n=168$. Note that the sample sizes are always the required number of patients for both of the two trial arms. Thus, to obtain a power of $80\%$, $309$ patients would have been required in both arms where equal randomization probabilities are assumed, which is much more than the recruited $119$ patients. As a consequence, even though the Bayes factor states moderate evidence for the null hypothesis $H_0:p_1=p_2$, our method shows that the trial had not sufficient patients included to achieve 80\% power to provide statistical evidence for $H_1:p_1<p_2$ with Bayes factors.\footnote{Even though the type-I-error is calibrated from a Bayesian perspective, the trial also did not recruit sufficient patients to achieve the desired 80\% probability of compelling evidence for $H_0$. Therefore, $168$ patients would have been required.}

This leads to the second question above, how large the power based on the $119$ recruited patients actually is. Even though the two arms are not balanced (which is assumed in \Cref{fig:riociguat_flat} for the power calculations), it is straightforward to calculate the power with our method for unbalanced trial arms. The methodology detailed in \Cref{sec:methodology} by no means needs to assume $n_1=n_2$, and for the $n=59$ patients in the treatment arm and the $n=60$ patients in the control arm, we set $n_1=60$, $n_2=59$, $y_1=38$ and $y_2=48$ and compute the associated (Bayesian) power. The latter results in $71.04\%$, which is below the desired $80\%$. We can also compute the type-I-error rate (Bayesian), which results in $0.017<0.05$ for the $119$ patients in the riociguat trial.

Now, the power and type-I-error rate in \Cref{fig:riociguat_flat} are computed as the Bayesian versions in \Cref{eq:bayesianDesign}. A regulatory agency like the FDA or EMA might require strict frequentist power and, in particular, type-I-error control. Therefore, it is reasonable to investigate the sample sizes needed also under these scenarios. To compute the frequentist type-I-error rate, it suffices to note that under $H_0:p_1=p_2$, the null set is the main diagonal in the unit cube $[0,1]$. Therefore, using a fine enough grid and computing the maximum of the type-I-error rate over all parameter combinations $(p_1,p_2)$ in this grid approximates the supremum to reject $H_0$ under this set, that is, the frequentist type-I-error rate.\footnote{For the directional test of $H_0:p_1=p_2$ versus $H_-:p_2 \leq p_1$ or $H_+:p_2 > p_1$ the computation works identical, while for the test of $H_-:p_2 \leq p_1$ versus $H_+:p_2 > p_1$ the null set $\{p_2 \leq p_1|p_1,p_2 \in [0,1]\}$ is a rectangle in the unit cube $[0,1]$. Using a grid over this rectangle approximates the supremum likewise then.} 
The maximum found via the grid-approximation results in a type-I-error of $0.09 > 0.05$. Thus, from a fully frequentist perspective, the design is not calibrated. Using a stricter evidence threshold $k$ such as $k=1/10$ can improve the frequentist type-I-error rate, but increases the number of patients to achieve the desired power simultaneously.

\begin{figure}[h!]
    \centering
    \includegraphics[width=0.9\linewidth]{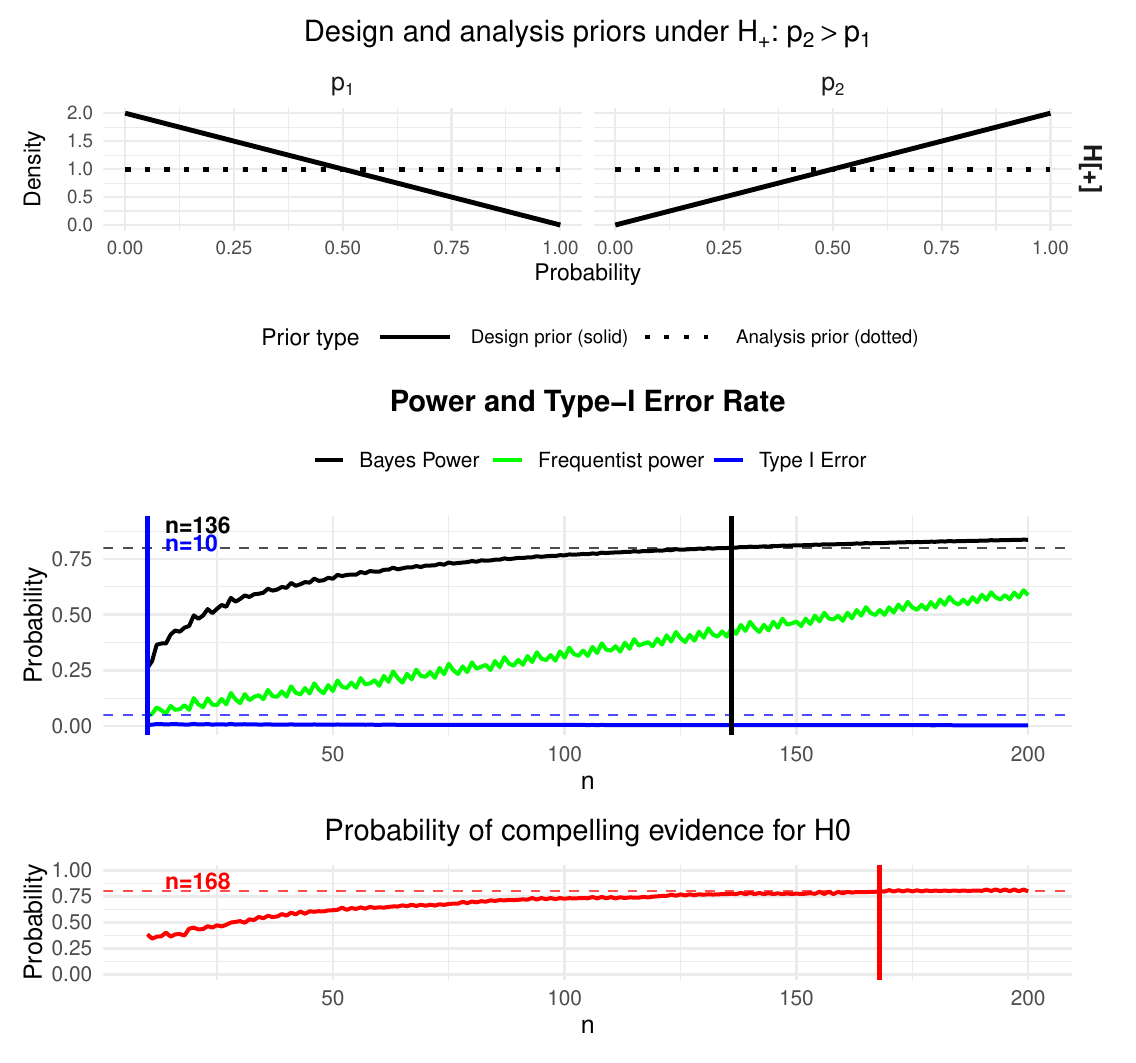}
    \caption{Bayesian power and sample size calculations for the riociguat trial, where $H_0:p_1=p_2$ versus $H_1:p_2>p_1$ is tested. Informative $B(1,2)$ and $B(2,1)$ design and flat analysis priors are used with strong evidence threshold $k=1/10$ and evidence threshold for compelling evidence under $H_0$ of $k_f=3$. The calibration shows the results for 80\% power, 5\% type-I-error rate and 80\% probability of compelling evidence under $H_0$. Frequentist power was obtained under $p_1=0.4$ and $p_2=0.6$.}
    \label{fig:riociguat_informative_strong}
\end{figure}

Therefore, \Cref{fig:riociguat_informative_strong} shows the results of our method when shifting to the evidence threshold for strong evidence, $k=1/10$, and using slightly more informative design priors. In \Cref{fig:riociguat_flat}, the design priors were flat and uninformative. However, we assume a priori that the proportion of patients with no progression will be smaller in the riociguat treatment arm than in the placebo group. Therefore, it is reasonable to assign smaller probabilities close to zero a higher a priori density value than large probabilities close to one, in the control arm. In the treatment arm, we would assign larger probabilities closer to one higher density values than smaller probabilities close to zero, expressing our expectation that the drug is effective. We therefore use $a_1^d=1, b_1^d = 2$, and $a_2^d = 2, b_2^d = 1$, shown in the top plot of \Cref{fig:riociguat_informative_strong}. Now, the middle plot in \Cref{fig:riociguat_informative_strong} shows that the required sample size is roughly halved for achieving 80\% Bayesian power: Only $n=136$ patients -- that is, $68$ per trial arm -- are required now. Also, the frequentist type-I-error rate is now calibrated and the maximum found over the null set results in $0.021<0.05$. The other metrics such as the Bayesian type-I-error rate and the probability of compelling evidence under $H_0$ do not change, as the design priors under $H_+$ are only associated with the Bayesian power.\footnote{It is, of course, possible to choose an informative design prior under $H_0$, that is, choose different values for $a_0^d$ and $b_0^d$. For example, it might be more realistic to assume that $p_1=p_2$ holds in the region $[0.25,0.75]$ than outside of it. Centering the design prior around $p=0.5$ would be a possibility to express this a priori beliefs under $H_0$.}

As a last point, consider the green line in the middle plots in \Cref{fig:riociguat_flat} and \Cref{fig:riociguat_informative_strong}. A frequentist power might also be warranted by a regulatory agency. Thus, we allow to specify a parameter combination $(p_1,p_2)$, expressing the expected effect between the treatment and control arm, and compute the frequentist power for the Bayes factor under these fixed parameter choices. In \Cref{fig:riociguat_flat} and \Cref{fig:riociguat_informative_strong}, we chose $p_1=0.4$ and $p_2=0.6$, expressing our expectation that in the placebo group 40\% of the patients are progression-free. In the treatment group, we assume this proportion increases further to about 60\%.\footnote{Note that from \Cref{tab:riociguat-mrss}, the trial data show that the true proportions are even larger. About $\approx 80\%$ of the riociguat arm do not experience progression, while in the control arm this proportion is about $\approx 66\%$. However, when planning a trial, this information often is not available or only available through historical information or other trial data. We thus recommend computing frequentist power under multiple realistic parameter choices for $p_1$ and $p_2$.} \Cref{fig:riociguat_flat} shows that in this case, $n=102$ patients per arm or $204$ in total are required to achieve the desired 80\% power in a fully frequentist sense. When shifting to \Cref{fig:riociguat_informative_strong}, the design priors do not influence the frequentist power calculation, but the stronger evidence threshold $k=1/10$ does. Therefore, the required number of patients increases in the middle plot of \Cref{fig:riociguat_informative_strong}, and even for $n=200$ patients in total the frequentist power requirement is not reached.

Taking stock, the moderate evidence in favour of $H_+$ indicated by the Bayes factor in the riociguat trial is not backed by solid long-term error guarantees. Under flat or informative design priors, whether a moderate or strong evidence threshold is applied, more than the recruited $n=119$ patients would have been required to achieve a sufficiently large (Bayesian) or frequentist power for $H_1$ and a sufficiently large probability of compelling evidence for $H_0$.

Our method clarifies that under slightly informative design priors, $136$ patients would have been required in total to yield a fully calibrated design, except for frequentist calibration of power when assuming $p_1=0.4$ and $p_2=0.6$, when a strong evidence threshold is used.

\subsubsection{ICT-107 Trial}\label{subsubsec:example2}
The ICT-107 phase II trial randomized 124 newly diagnosed glioblastoma patients (81 ICT-107 vaccine, 43 placebo) to assess immunologic response \citep{wenRandomizedDoubleBlindPlaceboControlled2019}. The ICT‑107 program is an autologous dendritic‑cell (DC) vaccine approach tested in newly diagnosed glioblastoma, with the key randomized evidence coming from a phase II, double‑blind, placebo‑controlled trial in 124 patients. Supplementary tables report exact binary immunologic responders:

\begin{table}[h]
\centering
\caption{Binary endpoint: immunologic response (ICT-107 trial)}
\begin{tabular}{|l|c|c|}
\hline
 & Responders (Success) & Non-responders \\
\hline
ICT-107 (n=81) & 49 & 32 \\
\hline
Placebo (n=43) & 12 & 31 \\
\hline
\end{tabular}
\label{tab:ict107}
\end{table}

These counts -- see \Cref{tab:ict107} -- (60.5\% vs 27.9\%) indicate a treatment benefit (OR $0.29$, $p<0.001$) \citep{wenRandomizedDoubleBlindPlaceboControlled2019}.

The trial tested superiority of ICT-107 vaccine, resembling $H_-:p_2\leq p_1$ vs. $H_+:p_2> p_1$. For the endpoint immunologic response, this implies that evidence for $H_+$ means treatment with ICT-107 leads to a higher proportion of patients with immunologic response (success) than placebo. Thus, ideally trial data provides evidence for $H_+$.

\begin{figure}[h!]
    \centering
    \includegraphics[width=0.9\linewidth]{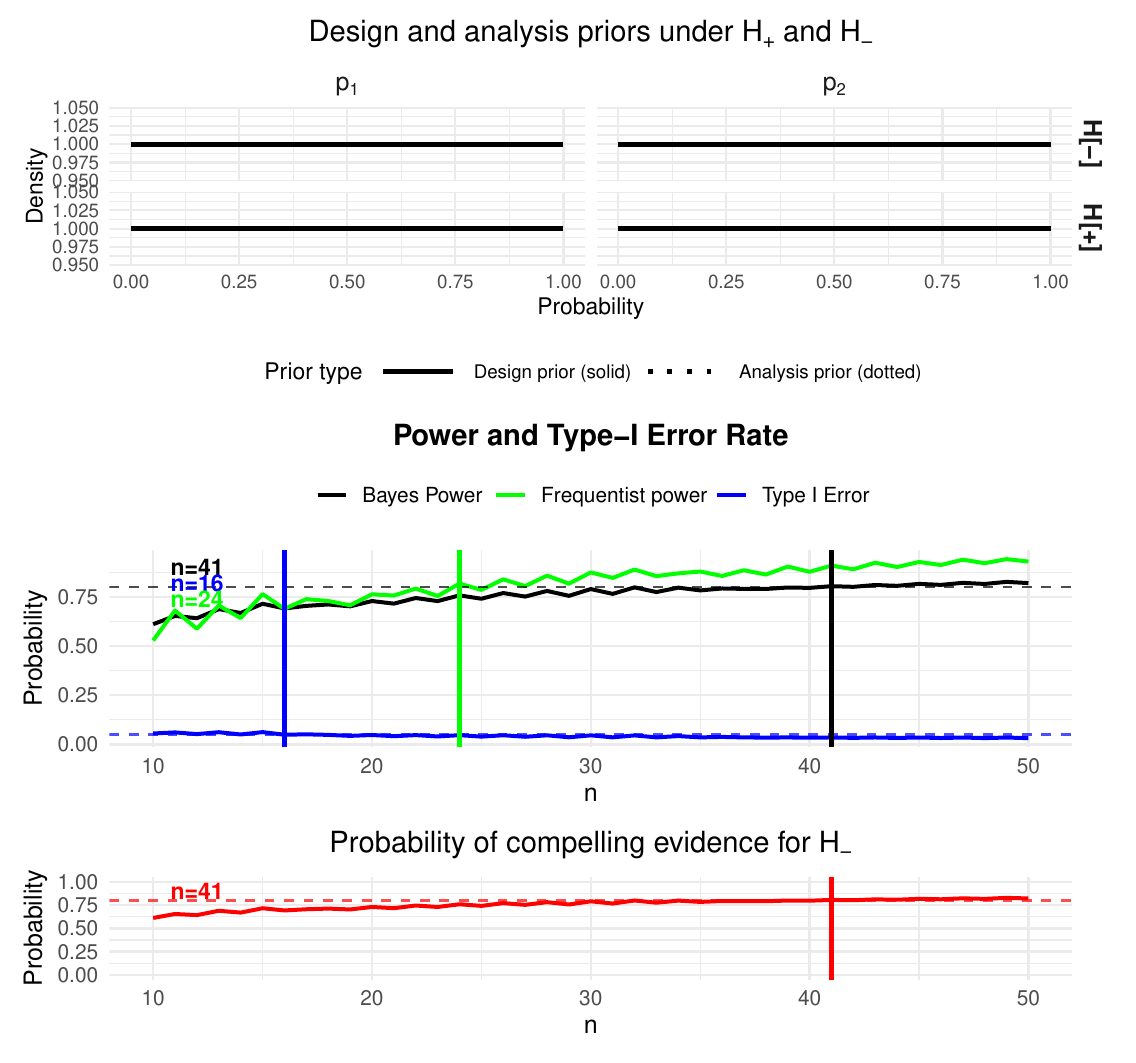}
    \caption{Bayesian power and sample size calculations for the ICT-107 trial, where $H_-:p_1\geq p_2$ versus $H_+:p_2>p_1$ is tested. Flat design and analysis priors are used with moderate evidence threshold $k=1/3$ and evidence threshold $k_f=3$ for compelling evidence for $H_-$. The calibration shows the results for 80\% power, 5\% type-I-error rate and 80\% probability of compelling evidence under $H_0$. Frequentist power was obtained under $p_1=0.3$ and $p_2=0.6$.}
    \label{fig:ex2_flat_moderate}
\end{figure}

We use the theory derived in \Cref{subsec:one-sided-hyp1} and use moderate thresholds both for the Bayesian power and the probability of compelling evidence under $H_-$, that is, $k=3$ and $k_f=3$ in \Cref{eq:bayesianDesign} and \Cref{eq:pce_H0}.

We use independent uniform analysis Beta$(1,1)$ priors for each arm in the analysis, both under $H_-$ and $H_+$. Under these priors the Bayes factor $\mathrm{BF}_{+-}$ in favor of $H_+$ results in $\mathrm{BF}_{+-}=3702.65$, indicating extreme evidence in favour of $H_+:p_2 >p_1$. Therefore, one would be tempted to conclude that immunologic response rates are larger under the vaccine. Still, it remains open whether the trial had sufficient power to find compelling evidence for $H_-$ if the latter is true. Also, it remains open which sample sizes would be needed to have a (Bayesian or frequentist) power of at least 80\% for $H_+$, and which sample sizes are required for a type-I-error control (e.g. 5\%). \Cref{fig:ex2_flat_moderate} provides answers to these questions when using flat design and analysis priors in both arms, both under $H_-$ and $H_+$. Assuming balanced randmization, the middle plot in \Cref{fig:ex2_flat_moderate} shows that $n=41$ patients in total suffice to achieve 80\% Bayesian power, so $n=21$ patients per trial arm are enough to assert 80\% Bayesian powr. Frequentist power is calibrated already for $n=24$ patients in total, where the latter assumes true proportions of $p_1=0.3$ and $p_2=0.5$ in the control and treatment arms. Thus, $12$ patients per trial arm suffice for 80\% frequentist power. The Bayesian type-I-error rate -- compare the vertical blue line in the middle plot in \Cref{fig:ex2_flat_moderate} -- is calibrated for $n=16$ patients in total, so $8$ patients per trial arm are enough. The probability of compelling evidence (shown in the bottom plot in \Cref{fig:ex2_flat_moderate}) requires $n=41$ patients in total, which is met by the original trial sample sizes, compare \Cref{tab:ict107}.\footnote{The frequentist type-I-error rate is provided in the numerical output of the \texttt{powerbinbf01()} function that is implemented in our R package \texttt{bfbin2arm}. We provide details how to use the function in the Appendix.} The Bayesian type-I-error rate is already calibrated at 5\% when $n_1=n_2=8$, but the frequentist type-I-error rate (not shown in the plot) peaks at $32.7\%$, which is not tolerable from a fully frequentist perspective.

\begin{figure}[h!]
    \centering
    \includegraphics[width=0.9\linewidth]{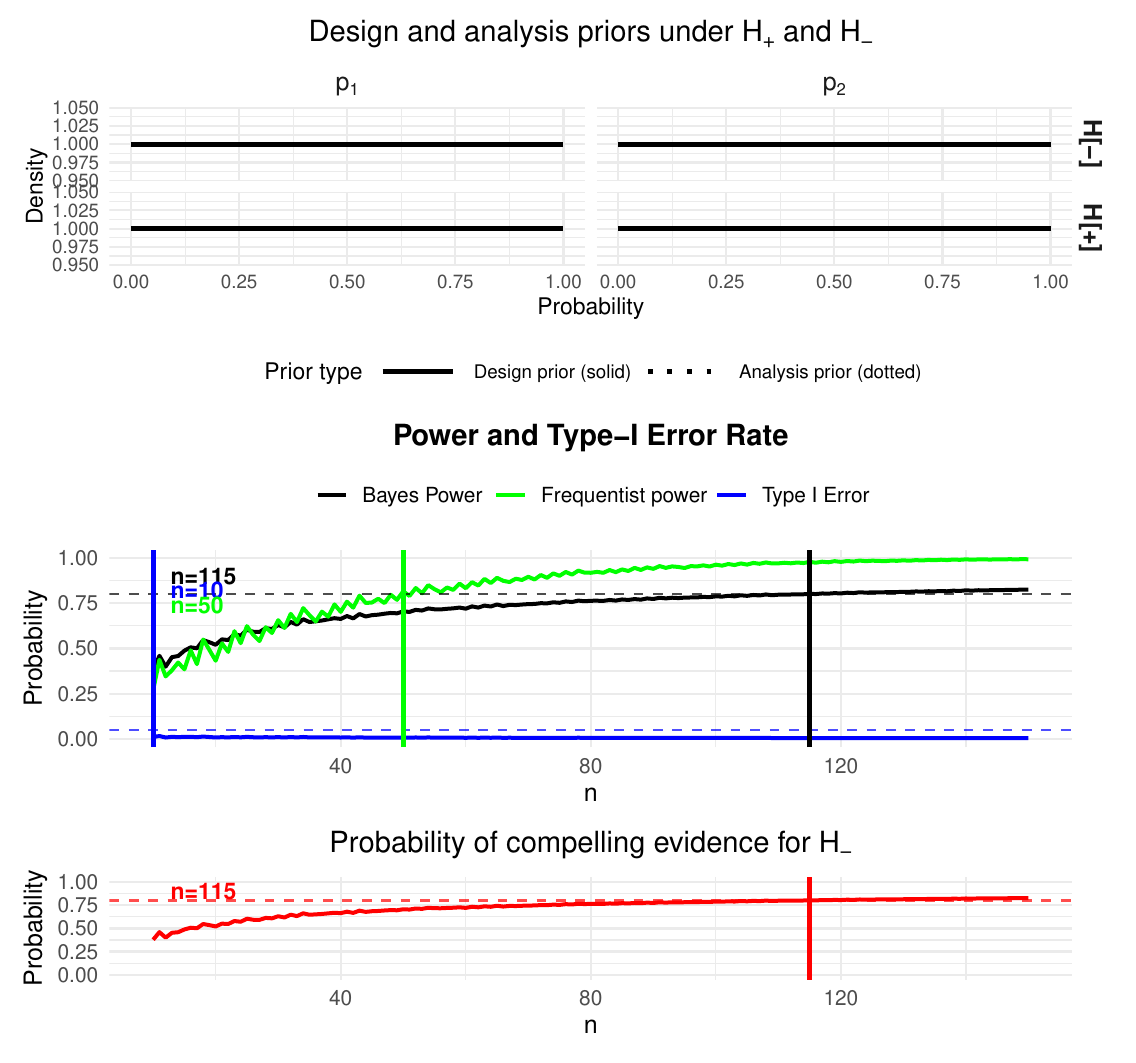}
    \caption{Bayesian power and sample size calculations for the ICT-107 trial, where $H_-:p_1\geq p_2$ versus $H_+:p_2>p_1$ is tested. Flat design and analysis priors are used with moderate evidence threshold $k=1/10$ and evidence threshold $k_f=10$ for compelling evidence for $H_-$. The calibration shows the results for 80\% power, 5\% type-I-error rate and 80\% probability of compelling evidence under $H_0$. Frequentist power was obtained under $p_1=0.3$ and $p_2=0.6$.}
    \label{fig:ex2_flat_strong}
\end{figure}

Therefore, \Cref{fig:ex2_flat_strong} provides the results when shifting to a strong evidence threshold $k=1/10$ instead of $k=1/3$. Now the (frequentist) type-I-error rate decreases to $12\%$, which is about a third of the one obtained when using $k=1/3$, but still above the desired 5\%. \Cref{fig:ex2_flat_strong} also shows that now more patients ($n=115$ in total in both arms for 80\% Bayesian power and $n=50$ for 80\% frequentist power in both arms) are required to achieve 80\% (Bayesian or frequentist) power, because the evidence threshold is more strict now. Now, based on these intermediate design results, two steps are necessary to provide a design which additionally is fully calibrated from a frequentist sense (5\% type-I-error): The type-I-error rate can be decreased further by shifting towards an even more strict evidence threshold.

\begin{figure}[h!]
    \centering
    \includegraphics[width=0.9\linewidth]{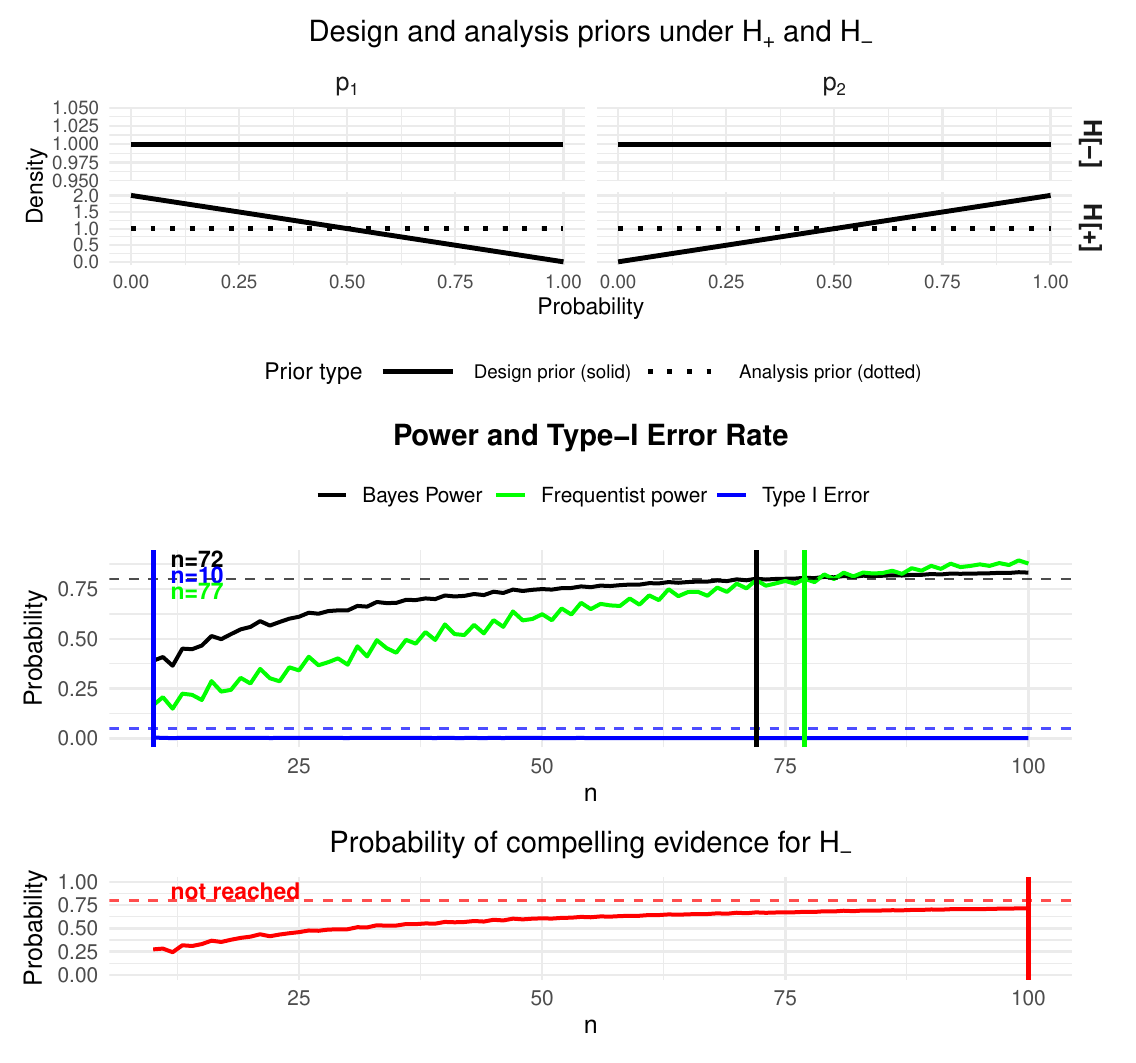}
    \caption{Bayesian power and sample size calculations for the ICT-107 trial, where $H_-:p_1\geq p_2$ versus $H_+:p_2>p_1$ is tested. Informative design priors $B(1,2)$ and $B(2,1)$ are used for the control and treatment group, and flat analysis priors are chosen. Results are based on a very strong evidence threshold $k=1/30$ and evidence threshold $k_f=30$ for compelling evidence for $H_-$. The calibration shows the results for 80\% power, 5\% type-I-error rate and 80\% probability of compelling evidence under $H_0$. Frequentist power was obtained under $p_1=0.3$ and $p_2=0.6$.}
    \label{fig:ex2_informative_verystrong}
\end{figure}

\Cref{fig:ex2_informative_verystrong} shows the results when shifting to an even stricter evidence threshold $k=1/30$ (and $k_f=30$), which amounts to very strong evidence according to the scale of \cite{Jeffreys1939}. As a consequence, the required sample size to achieve the desired 80\% power should increase. However, as the resulting $n=115$ patients per trial arm obtained under flat design priors and $k=1/10$ in \Cref{fig:ex2_flat_strong} yield a trial sample size already close to the $n=124$ patients used in the ICT-107 trial -- compare \Cref{tab:ict107} -- we also shift to slightly more informative design priors.\footnote{Importantly, the analysis priors are still flat and the Bayes factor is calculated based on these noninformative analysis priors. The latter is important from a regulatory agency's perspective such as the EMA or FDA \citep{ema2022reflectionpaper,FDABayes2010}.} These are shown in the top plots of \Cref{fig:ex2_informative_verystrong} and resemble our expectation about the efficacy of the vaccine. Thus, the shape of the design prior of $p_1$ under $H_+:p_2 > p_1$ is reflecting our belief that smaller probabilities are more likely a priori for $p_1$ than for $p_2$ (compare the top plot in \Cref{fig:ex2_informative_verystrong}). Likewise, larger probabilities are more likely a priori for $p_2$ than for $p_1$ under $H_+$. The resulting sample sizes of the calibrated phase II design with Bayes factors are shown in the middle and bottom plot of \Cref{fig:ex2_informative_verystrong}. 

For 80\% Bayesian power, $n=36$ patients suffice per trial arm. For 80\% frequentist power, $n=77$ patients per trial arm suffice. The Bayesian type-I-error rate is calibrated for even $n=10$ patients in total, and the frequentist type-I-error rate peaks at $0.041$ under $H_-$ here, and also is calibrated.\footnote{This is again computed via a grid-approximation of the set under $H_-$, and for each point in this grid -- which is a combination of parameter values $p_1$ and $p_2$ in the unit cube $[0,1$ -- the type-I-error rate is computed for the Bayes factor with threshold $k$ under the analysis priors (flat). The maximum among all parameter values (or points in this grid) is then reported as the frequentist type-I-error rate.} However, the bottom plot in \Cref{fig:ex2_informative_verystrong} shows that even for $n=100$, the probability of compelling evidence for $H_0$ does not meet the requirement of 80\%.\footnote{For $n=100$ patients per trial arm, our approach yields a probability of compelling evidence for $H_0$ of $71.5\%$, slightly below the required $80\%$.} Now, the reason why the probability of compelling evidence for $H_0$ requires much more patients than the power or type-I-error rate is that under $H_-$ the design priors are still flat. This makes it difficult to accumulate evidence in favour of $H_-$ as no expectation about $p_1$ and $p_2$ and their relationship under $H_-$ is reflected in the shape of the design priors.

\begin{figure}[h!]
    \centering
    \includegraphics[width=0.9\linewidth]{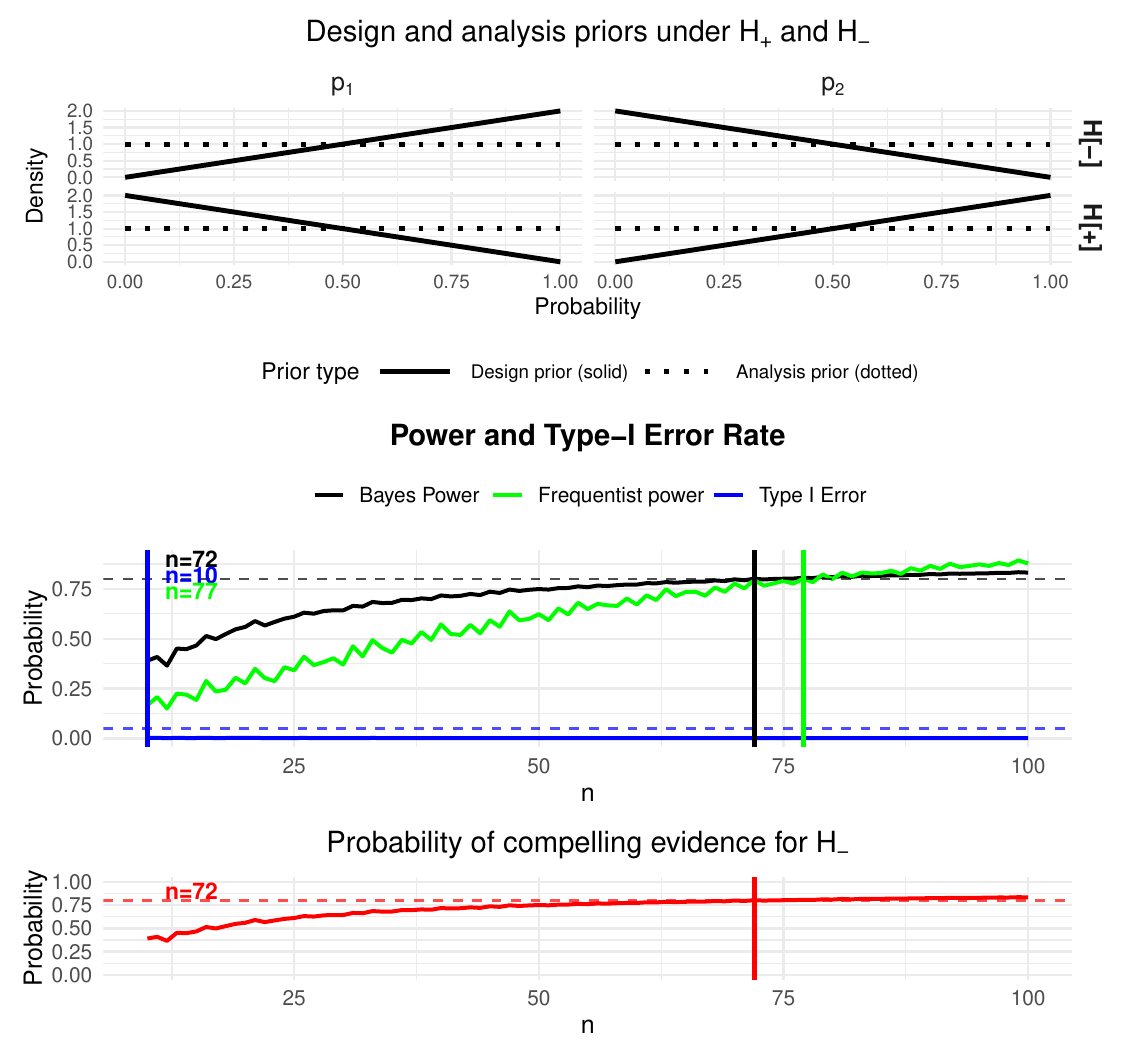}
    \caption{Bayesian power and sample size calculations for the ICT-107 trial, where $H_-:p_1\geq p_2$ versus $H_+:p_2>p_1$ is tested. Informative design priors $B(1,2)$ and $B(2,1)$ are used for the control and treatment group, and flat analysis priors are chosen. Additionally now, the design priors under $H_-:p_1 \geq p_2$ are also informative and chosen as $B(2,1)$ and $B(1,2)$ for the control and treatment group. Results are based on a very strong evidence threshold $k=1/30$ and $k_f=30$ for compelling evidence for $H_-$. The calibration shows the results for 80\% power, 5\% type-I-error rate and 80\% probability of compelling evidence under $H_0$. Frequentist power was obtained under $p_1=0.3$ and $p_2=0.6$.}
    \label{fig:ex2_informative_very_strong_Hminus}
\end{figure}

Thus, \Cref{fig:ex2_informative_very_strong_Hminus} shows the final results when also modifying the design priors under $H_-$ into a slightly more informative shape, reflecting that under $H_-:p_2 \leq p_1$ we believe the probabilities $p_1$ to be larger or equally large as $p_2$. \Cref{fig:ex2_informative_very_strong_Hminus} shows that the results for the (Bayesian or frequentist) power in the middle plot remain the same, as these only rely on the evidence threshold $k$ and the design priors under $H_+$. Also, the frequentist and Bayesian type-I-error rate results stay the same. The bottom plot shows that now, $n=72$ patients in total or $n=36$ patients per trial arm suffice to calibrate the probability of compelling evidence for $H_-$. Thus, we can summarize the final results in \Cref{fig:ex2_informative_very_strong_Hminus} as follows: When slightly informative design priors are chosen both under $H_+$ and $H_-$, the design is fully calibrated from a Bayesian point of view, regarding 80\% power and 5\% type-I-error rate (for $k=1/30$) when at least $n=36$ patients are in each trial arm. Furthermore, the probability of compelling evidence is calibrated at 80\% then, too (for $k=30$). From a fully frequentist point of view, the type-I-error rate is calibrated at 5\%, too, when $n=36$ patients are included in each trial arm. Importantly, assuming $p_1=0.3$ and $p_2=0.6$, the frequentist power is only calibrated at 80\% for $n=77$ patients in total, which implies $n=39$ patients per trial arm assuming balanced randomization.

Thus, taking stock, the ICT-107 trial, which yielded an extreme Bayes factor of $\mathrm{BF}_{+0}=3702.65$, indicating extreme evidence in favour of $H_+:p_2 >p_1$, is fully calibrated based on the sample sizes in \Cref{tab:ict107}. All relevant metrics are calibrated, in particular the Bayesian power, Bayesian and frequentist type-I-error. Additionally, the probability of compelling evidence requirement under $H_0$ is satisfied from a Bayesian point of view. This strengthens the already convincing result displayed in \Cref{tab:ict107}.\footnote{We acknowledge that the original trial randomization scheme did not use balanced randomization but randomized one third of the patients into the control while two-thirds were randomized into the treatment group. Our implementation in the \texttt{binbf2arm} R package allows to design and calibrate Bayes factor designs with unbalanced randomizations, and we provide the results for the original ICT-107 randomization with example code in the Appendix.}

\section{Discussion}\label{sec:discussion}

This paper introduces a simulation-free, numerically exact approach for Bayesian power and sample size calculations using Bayes factors in two-arm clinical trials with binary endpoints. Our method addresses a critical gap in Bayesian trial design by providing computationally efficient solutions that calibrate long-run error rates and power without Monte Carlo simulation or asymptotic approximations. The derivations for point-null versus composite ($H_0:p_1=p_2$ vs $H_1:p_1\neq p_2$) and directional hypotheses ($H_0:\eta=0$ vs $H_+:\eta > 0$, $H_0:\eta=0$ vs $H_-:\eta < 0$, $H_-:\eta \leq 0$ vs $H_+:\eta > 0$) enable precise planning and design for phase II trials.

\subsection{Key Advantages Over Simulation-Based Methods}

Traditional Bayesian sample size determination relies on Monte Carlo simulation, requiring thousands of iterations to estimate error rates with acceptable precision \citep{Morris2019,Siepe2024,Kelter2023,KelterPawel2025,PawelHeld2024,Seibold2021}. Our matrix-based root-finding approach (see Figure~\ref{fig:overview}) evaluates all possible data combinations exactly via prior-predictive densities and Beta functions, eliminating Monte Carlo standard errors and ensuring reproducibility. A call of our calibration function with in the range of $n=1$ to $n=100$ patients per arm completes in seconds versus hours of simulation on a modern desktop computer, with no loss of precision.

The $R$ package \texttt{bfbin2arm} implements all Bayes factors analytically, supporting both integer and non-integer prior shapes through finite-sum representations and numerical integration.\footnote{For guidance on how to use the package, see the Appendix.} Regulatory submission of simulation code is challenging due to random seeds and convergence diagnostics; our deterministic approach provides transparent, verifiable results.

\subsection{Real-World Calibration Insights}

 Our reanalysis of the riociguat and ICT-107 trials revealed common design challenges when using Bayes factors in phase II trials. Despite promising Bayes factors ($\mathrm{BF}_{+0}=4.32$, $\mathrm{BF}_{+-}=3702$), both trials lacked power guarantees for their sample sizes under noninformative priors. Flat design priors demand unrealistically large sample sizes in both cases, while slightly informative priors reflecting clinical expectations produce realistic sample sizes for fully calibrated designs. This phenomenon can be attributed to the geometry of the priors, compare \Cref{fig:priors}: Flat priors do not separate the hypotheses under consideration in any reasonable form, and thus the required sample sizes for strong or very strong evidence thresholds $k$ become large. Once slightly informative priors separate the hypotheses under comparision in a meaningful way, the required sample size drops quickly to realistic sample sizes for a phase II trial with two arms. This is, while still using flat analysis priors and being in concordance with regulatory guidelines of the Food and Drug Administration (FDA) and European Medicines Agencies (EMA) (for details see the next subsection). 

Directional testing proved superior for superiority claims in our re-analysis: $\mathrm{BF}_{+-}$ for $H_-:\eta\leq 0$ vs $H_+:\eta>0$ requires $n=36$ per arm (slightly informative design priors, $k=1/30$) with full Bayesian calibration (80\% power, $\leq 5\%$ Bayesian type-I-error rate, 80\% probability of compelling evidence $P(\mathrm{BF}_{10}>30|H_-)$) and frequentist Type I error peaking at 4\%. ICT-107's 124 patients thus provided robust evidence beyond its frequentist analysis (OR=0.29, $p<0.001$).

\subsection{Regulatory Alignment}

The FDA and EMA guidance accepts Bayesian methods which demonstrate ``sufficiently robust'' frequentist characteristics \citep{FDABayes2010,ema2022reflectionpaper}. Our calibration exceeds these requirements: Bayesian Type I error $\leq5\%$ aligns with $\alpha=0.05$, while the frequentist type-I-error calculation over null regions provides strict frequentist type-I-error control. The option to calibrate the probability of compelling evidence for $H_0$, $P(\mathrm{BF}_{01}>k_f|H_0)\geq80\%$ ensures futility detection in case a treatment or drug is not working, which often is absent in fixed-threshold posterior probability designs. The official Guidance for the Use of Bayesian Statistics in Medical Device Clinical Trials released by the FDA states that
\begin{quote}
    \textit{''Pure'' Bayesian approaches to statistics do not necessarily place the same emphasis on the notion of control of type I error as traditional frequentist approaches. There have, however, been some proposals in the literature that Bayesian methods should be ``calibrated'' to have good frequentist properties (e.g. Rubin, 1984; Box, 1980). In this spirit, as well as in adherence to regulatory practice, FDA recommends you provide the type I and II error rates of your proposed Bayesian analysis plan.} \citep[p.~29]{FDABayes2010}
\end{quote}
Likewise, the EMA requires sponsors to show appropriate type-I error control and adequate power when Bayesian methods underpin primary inferences \citep[p.~24]{europeanmedicinesagencyICHE20Adaptive2025}.

Given the broad range of options to calibrate a Bayesian design by our proposed method, the developed approach is in alignment with regulatory recommendations. Furthermore, Jeffreys' scale thresholds ($k=1/3,1/10,1/30$) yield interpretable evidence levels (moderate/strong/very strong) with sample sizes that are realistic for phase II trials. The Bayes-frequentist compromise satisfies regulators preferring objective analysis priors while allowing informative design priors to incorporate prior beliefs about the treatment efficacy during the planning stage.

Recentl, the U.S. Food and Drug Adniminstration released a draft on the use of Bayesian Methodology in Clinical Trials of Drug and Biological Products, and emphasized the following:

\begin{quote}
\textit{For (...) Bayesian approaches, specification of a success criterion is most often based on the posterior probability that the true treatment effect size exceeds some threshold. In mathematical notation, such a criterion might take the form $Pr(d > a) > c$, where $d$ is a population-level summary of the size of the treatment effect, $a$ is a minimum threshold for the treatment effect to be considered beneficial, and $c$ is a minimum probability level that would support a conclusion of effectiveness. (...) Choice of a success criterion of this kind thus means choice of specific values for $a$ and for $c$. There are a variety of approaches to specifying these thresholds for Bayesian analyses. The choice of which approach to use depends on the trial objectives and specific Bayesian methods used.} \citep[p.~6]{FDA_UseOfBayesianMethodologyJanuary2026}
\end{quote}
In contrast to the posterior probability, our design makes use of Bayes factors as the success criterion. Two comments are worth mentioning here:
\begin{enumerate}
    \item[$\blacktriangleright$]{First, our design can be extended to the use of posterior probability if desired, as the posterior probability is simply the Bayes factor multiplied by the prior odds of the hypotheses under consideration. While this is not currently implemented in the \texttt{bfbin2arm} R package, an extension is straightforward.}
    \item[$\blacktriangleright$]{Second, the form of our hypotheses differs in the sense that e.g. for $H_+:p_2-p_1>0$ versus $H_-:p_2-p_1\leq 0$ the minimum threshold $a$ above equals zero, while $p_2-p_1$ is the population-level summary of the size of the treatment effect. Different values of $a$ require different hypotheses, e.g. $H_+:p_2-p_1>0.1$ versus $H_-:p_2-p_1\leq 0.1$. This, in turn, requires a modification of the truncated priors used under both hypotheses, compare \Cref{fig:priors}. However, such modifications are also straightforward. For example, for the hypotheses $H_+:p_2-p_1>0.1$ versus $H_-:p_2-p_1\leq 0.1$ the truncated priors are simply the ones restricted not by the main diagonal in the two-dimensional unit-square, but the ones restricted by this very diagonal shifted up by $0.1$ on the $p_2$-axis in \Cref{fig:priors}.}
\end{enumerate}
Thus, this shows that our design is in line with the current thoughts of the FDA on the use of Bayesian Methodology in Clinical Trials. Whether a modification of the hypotheses as mentioned in the second point above is necessary, is up the the users, as ``when calibrating Bayesian success criteria to Type I error rate, $a$ is chosen to be 0 for superiority designs'' \citep[p.~6]{FDA_UseOfBayesianMethodologyJanuary2026}. Furthermore, when it comes to type-I-error calibration, the draft stresses that
\begin{quote}
    \textit{For some Bayesian approaches, $a$ and $c$ can be chosen such that the overall FWER is controlled at a given level, typically 0.025 one-sided. This is referred to as calibrating the success criteria to Type I error rate. Such an approach may be appropriate for designs where Bayesian approaches are used not to synthesize multiple information sources, but instead to facilitate complex adaptive designs. Calibration to Type I error rate also may be useful in designs with noninformative prior distributions that express a lack of prior information relevant to the analysis.} \citep[p.~6]{FDA_UseOfBayesianMethodologyJanuary2026}
\end{quote}
which is precisely in line with the approach used in our design. A distinctive feature of our design is the avoidance of any use of Monte Carlo simulations to calibrate Bayesian results to type-I-error rates. The FDA draft mentions that this is still the dominant approach in practice right now:
\begin{quote}
    \textit{For trial designs that calibrate Bayesian results to Type I error rate, the primary operating characteristics are the same as those described above for trials with frequentist inference (...). Clinical trial simulations are generally used to estimate or demonstrate control of operating characteristics. Briefly, a large number of simulated trials, conditional on a chosen prior distribution and sample size, are generated under the assumption that the null hypothesis is true or that an alternative hypothesis is true. The proportions of simulated trials in which the null hypothesis is rejected is then used to estimate Type I error rate and power, respectively.} \citep[p.~8]{FDA_UseOfBayesianMethodologyJanuary2026}
\end{quote}
Our novel calibration approach, in contrast, solely makes use of numerical methods, avoiding the computationally intensive simulations and need to report Monte Carlo standard errors, compare \cite{Kelter2023} and \cite{Seibold2021}. It is important to stress that that while a simulation in principle is possible, our approach yields exact results with substantially less computational effort.

Similar requirements can be found in the draft of the FDA on Complex and Innovative Trial Designs, which recommends that \textit{``detailed evaluation of the operating characteristics of the design, including its chance of producing erroneous conclusions and the reliability of treatment effect estimates. Type I error probability control and power should be addressed where applicable.''} \citep[p.~8]{FDA_ComplexInnovativeDesignsDecember2020}, while the key method to analyze this operating characteristics is by means of simulation-based methods.

Taking stock, our approach is fully in line with current regulatory recommendations, and even circumvents the use of computationally costly simulations and their associated problems with sponsors and clinicians like the implementation, communication and reproducibility of a trial design.

\subsection{Limitations and Extensions}

The discrete binomial model induces step-function behavior in power curves due to lattice effects, which is visible in all figures in this paper. Our implemented 10-sample lookahead approach in the \texttt{binbf2arm} ensures stability and worked without any problems in all examples, so users can be reasonably certain that the power does not drop below a specified threshold for the next ten sample sizes (likewise for the type-I-error rate and probability of compelling evidence). One main limitation of our method is that fixed sample sizes are assumed, which preclude interim analysis. However, sequential extensions follow naturally from recursive prior-predictives \citep{KelterPawel2025} and can be tackled in future work. For single-arm trials with binary endpoints, \cite{kelterBayesianOptimalTwostage2025} recently developed an efficient numerical approach to conduct two-stage sequential Bayesian trial design, recovering Simon's two-stage design as a special case in a variety of settings. The approach detailed there can be generalized also to the two-arm setting. Unequal randomization ($n_1\neq n_2$) can be handled by the implementation in the \texttt{binbf2arm} package, but assumes known allocation ratios.



This framework bridges Bayesian evidence quantification with frequentist operating characteristics, facilitating regulatory acceptance of efficient phase II designs in resource-constrained settings via the use of Bayes factors.

\section*{Acknowledgements}
The author is grateful to Samuel Pawel for feedback on the methodology and Silke Jörgens for helpful recommendations on literature on the positions of the official regulatory agencies regarding the use of Bayesian methodology in clinical trials.

\appendix
\appendixpage

\section{The \texttt{bfbin2arm} R package}
The \texttt{bfbin2arm} package is available on CRAN (\url{https://cran.r-project.org/web/packages/bfbin2arm/index.html}) and enables one-line calibration of a two-arm phase II clinical trial design using Bayes factors:
\begin{verbatim}
install.packages("bfbin2arm"); # install the package from CRAN
library(bfbin2arm); # load the package
ntwoarmbinbf01(power=0.8, alpha=0.05, pce_H0=0.8, 
               test="BF+-", k=1/10, nrange=c(20,200))
\end{verbatim}
The re-analysis of the ICT-107 trial is possible for example using the following set of parameters for the `ntwoarmbinbf01` function:
\begin{verbatim}
ntwoarmbinbf01(   
   k = 1/30, k_f = 30,
   power = 0.8, alpha = 0.05, pce_H0 = 0.8,
   test = "BF+-",
   nrange = c(10, 100), n_step = 1,
   progress = TRUE,
   a_1_d = 1, b_1_d = 2,
   a_2_d = 2, b_2_d = 1,
   a_1_d_Hminus = 2, b_1_d_Hminus = 1,
   a_2_d_Hminus = 1, b_2_d_Hminus = 2,
   compute_freq_t1e = TRUE,
   p1_power = 0.3, p2_power = 0.6,
   output = "plot"  # Returns recommended n per group
 )
\end{verbatim}
For details and further explanations on the use of the package see the provided vignette on CRAN. For illustration purposes, we showcase how to design and calibrate the ICT-107 trial discussed in detail in \Cref{subsubsec:example2} using the original randomization scheme, where one-third of the patients is randomized into the control group and two-thirds are randomized into the treatment group.

\begin{figure}[h!]
    \centering
    \includegraphics[width=1\linewidth]{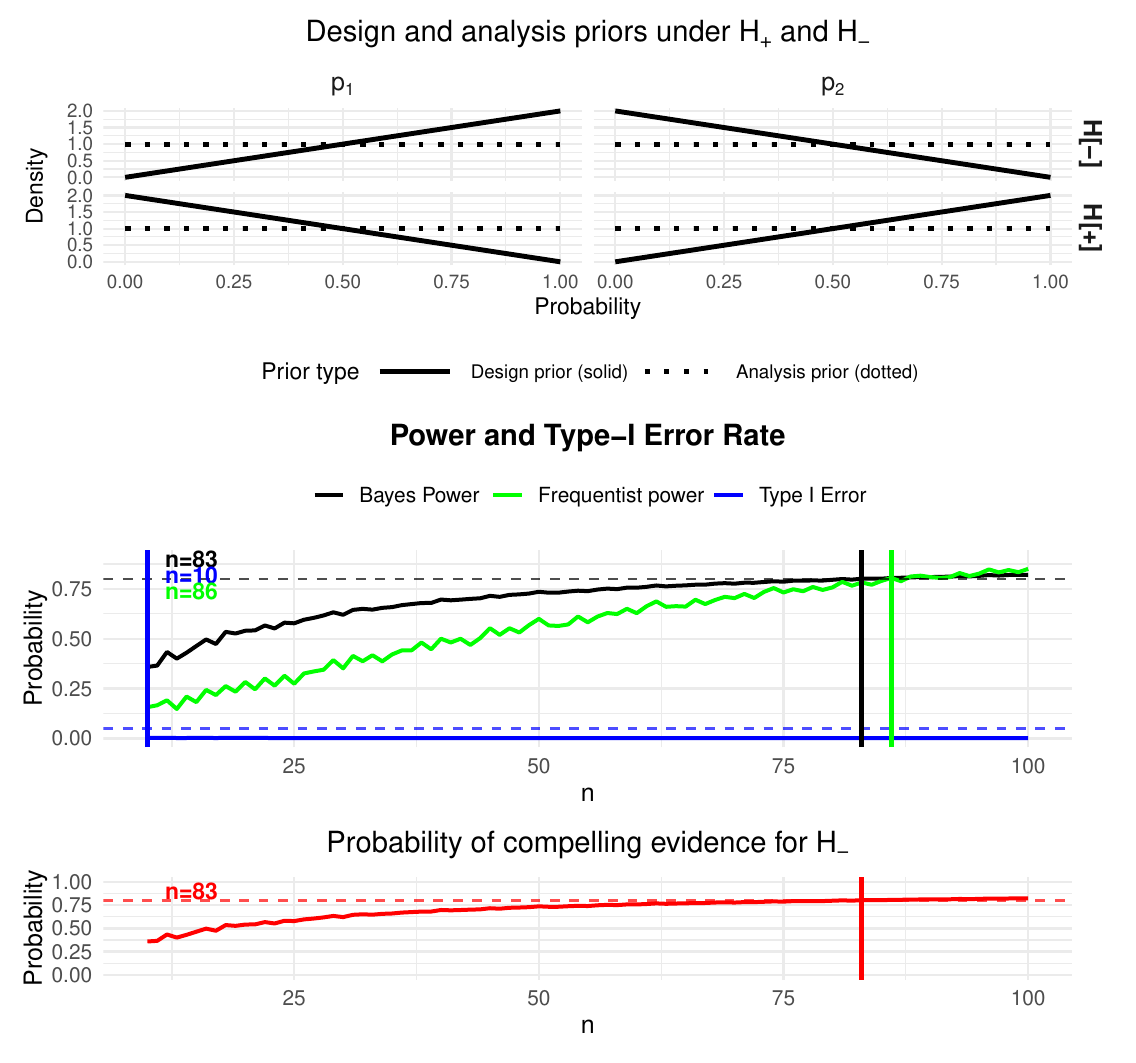}
    \caption{Bayesian power and sample size calculations for the ICT-107 trial, where $H_-:p_1\geq p_2$ versus $H_+:p_2>p_1$ is tested. Informative design priors $B(1,2)$ and $B(2,1)$ are used for the control and treatment group, and flat analysis priors are chosen. Additionally now, the design priors under $H_-:p_1 \geq p_2$ are also informative and chosen as $B(2,1)$ and $B(1,2)$ for the control and treatment group. Results are based on a very strong evidence threshold $k=1/30$ and $k_f=30$ for compelling evidence for $H_-$. The calibration shows the results for 80\% power, 5\% type-I-error rate and 80\% probability of compelling evidence under $H_0$. Frequentist power was obtained under $p_1=0.3$ and $p_2=0.6$. The randomization probabilities are $1/3$ for the control and $2/3$ for the treatment arm.}
    \label{fig:ex2_informative_very_strong_originalRandomization}
\end{figure}

\begin{verbatim}
ntwoarmbinbf01(   
   k = 1/30, k_f = 30,
   power = 0.8, alpha = 0.05, pce_H0 = 0.8,
   test = "BF+-",
   nrange = c(10, 100), n_step = 1,
   progress = TRUE,
   a_1_d = 1, b_1_d = 2,
   a_2_d = 2, b_2_d = 1,
   a_1_d_Hminus = 2, b_1_d_Hminus = 1,
   a_2_d_Hminus = 1, b_2_d_Hminus = 2,
   compute_freq_t1e = TRUE,
   p1_power = 0.3, p2_power = 0.6,
   output = "plot",
   alloc1 = 1/3,
   alloc2 = 2/3
 )
\end{verbatim}
Additionally to the former function call to \texttt{ntwoarmbinbf01}, now the \texttt{alloc1} and \texttt{alloc2} parameters are provided, specifying the randomization scheme. \texttt{alloc1} is the randomization probability for the control group and \texttt{alloc2} the one for the treatment group. \Cref{fig:ex2_informative_very_strong_originalRandomization} shows the output produced by the function call and shows that $n=83$ patients in total suffice to provide 80\% Bayesian power. This implies that $1/3\cdot 83 \approx 28$ patients in the control arm and $2/3 \cdot 83 \approx 56$ patients in the treatment arm are required to provide 80\% Bayesian power, which is met by the original ICT-107 sample sizes, compare \Cref{tab:ict107}.

\section{Derivations for the two-sided hypothesis test of $H_0:\eta = 0$ versus $H_1:\eta \neq 0$}

\begin{proof}[Prior-predictive under $H_1:\eta \neq 0$]
    \begin{align}
    f(d|H_0)&=f(y_1,y_2 \mid n_1, n_2, a_0^d, b_0^d)\nonumber\\
    &=\int \mathrm{Bin}(y_1|n_1,p)\mathrm{Bin}(y_2|n_2,p)\mathrm{Beta}(p|a_0^d,b_0^d)dp\nonumber\\
    &=\int {n_1 \choose y_1}p^{y_1}(1-p)^{n_1-y_1}{n_2 \choose y_2}p^{y_2}(1-p)^{n_2-y_2}\frac{p^{a_0^d-1}(1-p)^{b_0^d-1}}{\mathrm{B}(a_0^d,b_0^d)}dp\nonumber\\
    &=\frac{{n_1 \choose y_1}{n_2 \choose y_2}}{\mathrm{B}(a_0^d,b_0^d)} \int p^{y_1+y_2+a_0^d-1}(1-p)^{n_1+n_2-y_1-y_2+b_0^d-1}dp\nonumber\\
    &=\frac{{n_1 \choose y_1}{n_2 \choose y_2}}{\mathrm{B}(a_0^d,b_0^d)} \mathrm{B}(a_0^d+y_1+y_2,b_0^d+n_1+n_2-y_1-y_2)
\end{align}
which is \Cref{eq:predDensityH0}.
\end{proof}

\begin{proof}[Prior-predictive under $H_0:\eta=0$]
    \begin{align}
        f(d|&H_1)=f(y_1,y_2 \mid n_1, n_2, a_1^d, b_1^d, a_2^d, b_2^d)\nonumber\\
        &=\int \int \mathrm{Bin}(y_1|n_1,p_1)\mathrm{Beta}(p_1|a_1^d,b_1^d)\mathrm{Bin}(y_2|n_2,p_2)\mathrm{Beta}(p_2|a_2^d,b_2^d)dp_1 dp_2\nonumber\\
        &=\int \left [\int \mathrm{Bin}(y_1|n_1,p_1)\mathrm{Beta}(p_1|a_1^d,b_1^d)dp_1 \right ] \mathrm{Bin}(y_2|n_2,p_2)\mathrm{Beta}(p_2|a_2^d,b_2^d) dp_2\nonumber\\
        &=\int \mathrm{Bin}(y_1|n_1,p_1)\mathrm{Beta}(p_1|a_1^d,b_1^d)dp_1 \cdot \int \mathrm{Bin}(y_2|n_2,p_2)\mathrm{Beta}(p_2|a_2^d,b_2^d) dp_2\nonumber\\
        &=\int {n_1 \choose y_1}\frac{p_1^{y_1}(1-p_1)^{n_1-y_1}p_1^{a_1^d-1}(1-p_1)^{b_1^d-1}}{\mathrm{B}(a_1^d,b_1^d)}dp_1 \nonumber\\
        &\cdot \int {n_2 \choose y_2} \frac{p_2^{y_2}(1-p_2)^{n_2-y_2}p_2^{a_2^d-2}(1-p_2)^{b_2^d-1}}{\mathrm{B}(a_2^d,b_2^d)}dp_2\nonumber\\
        &={n_1 \choose y_1}\frac{\mathrm{B(y_1+a_1^d,n_1-y_1+b_1^d)}}{\mathrm{B}(a_1^d,b_1^d)} \cdot  {n_2 \choose y_2} \frac{\mathrm{B(y_2+a_2^d,n_2-y_2+b_2^d)}}{\mathrm{B}(a_2^d,b_2^d)}
\end{align}
which is \Cref{eq:predDensityH1}.
\end{proof}

\begin{proof}[Finite-sum form for $C$ (integer shapes)]
Using the expansion (for integer \(a_1\))
\[
I_x(a_1,b_1)
= \frac{1}{B(a_1,b_1)}
\sum_{k=0}^{a_1-1}
\binom{a_1+b_1-1}{k}\,
x^{k}(1-x)^{a_1+b_1-1-k},
\]
we obtain
\begin{align}
C
&= \int_0^1
\frac{p_2^{a_2-1}(1-p_2)^{b_2-1}}{B(a_2,b_2)}\,
I_{p_2}(a_1,b_1)\,dp_2 \nonumber\\
&= \frac{1}{B(a_1,b_1)B(a_2,b_2)}
\sum_{k=0}^{a_1-1}
\binom{a_1+b_1-1}{k}
\int_0^1
p_2^{a_2-1+k}(1-p_2)^{b_2-1+a_1+b_1-1-k}\,dp_2 \nonumber\\
&= \frac{1}{B(a_1,b_1)B(a_2,b_2)}
\sum_{k=0}^{a_1-1}
\binom{a_1+b_1-1}{k}\,
B\bigl(a_2+k,\;b_2+a_1+b_1-1-k\bigr).\label{eq:finiteSumFormC}
\end{align}
\end{proof}

\section{Derivations for the two-sided hypothesis test of $H_0:\eta = 0$ versus $H_+:\eta > 0$}

\begin{proof}[Predictive density under $H_+:\eta >0$]
\[
p(y_1,y_2 \mid H_+)
= \iint_{A} f(p_1,p_2 \mid y_1,y_2)\,\pi_+(p_1,p_2)\,dp_1\,dp_2.
\]
Substituting the truncated prior,
\[
\pi_+(p_1,p_2)
= \frac{1}{C}\,
\frac{p_1^{a_1^d-1}(1-p_1)^{b_1^d-1}}{B(a_1^d,b_1^d)}
\frac{p_2^{a_2^d-1}(1-p_2)^{b_2^d-1}}{B(a_2^d,b_2^d)}
\,\mathbf{1}\{0<p_1<p_2<1\},
\]
gives
\[
\begin{aligned}
p(y_1,y_2 \mid H_+)
&= \binom{n_1}{y_1}\binom{n_2}{y_2}
\frac{1}{B(a_1^d,b_1^d)B(a_2^d,b_2^d)\,C}
\int_0^1\int_{p_1}^1
p_1^{y_1+a_1^d-1}(1-p_1)^{n_1-y_1+b_1^d-1} \\
&\qquad\qquad\qquad\qquad\quad
\times p_2^{y_2+a_2^d-1}(1-p_2)^{n_2-y_2+b_2^d-1}
\,dp_2\,dp_1.
\end{aligned}
\]
Define the updated shape parameters
\begin{align}
A_1 = y_1 + a_1^d,\quad B_1 = n_1 - y_1 + b_1^d,\qquad
A_2 = y_2 + a_2^d,\quad B_2 = n_2 - y_2 + b_2^d.
\end{align}
Then the inner integral, for fixed \(p_1\), is
\[
\int_{p_1}^1 p_2^{A_2-1}(1-p_2)^{B_2-1}\,dp_2
= B(A_2,B_2) - B_{p_1}(A_2,B_2)
= B(A_2,B_2)\bigl[1 - I_{p_1}(A_2,B_2)\bigr].
\]
Hence
\[
\begin{aligned}
p(y_1,y_2 \mid H_+)
&= \binom{n_1}{y_1}\binom{n_2}{y_2}
\frac{1}{B(a_1^d,b_1^d)B(a_2^d,b_2^d)\,C}
\int_0^1
p_1^{A_1-1}(1-p_1)^{B_1-1} \\
&\qquad\qquad\qquad\qquad\quad
\times B(A_2,B_2)\bigl[1 - I_{p_1}(A_2,B_2)\bigr]
\,dp_1.
\end{aligned}
\]
Define
\[
I(y_1,y_2)
= \int_0^1\int_{p_1}^1
p_1^{A_1-1}(1-p_1)^{B_1-1}
p_2^{A_2-1}(1-p_2)^{B_2-1}
\,dp_2\,dp_1,
\]
so
\begin{align}
    p(y_1,y_2 \mid H_+)
= \binom{n_1}{y_1}\binom{n_2}{y_2}
\frac{I(y_1,y_2)}{B(a_1^d,b_1^d)B(a_2^d,b_2^d)\,C}.
\end{align}
Using the expression for the inner integral,
\[
\begin{aligned}
I(y_1,y_2)
&= \int_0^1
p_1^{A_1-1}(1-p_1)^{B_1-1}
\bigl[B(A_2,B_2) - B_{p_1}(A_2,B_2)\bigr]
\,dp_1 \\
&= B(A_2,B_2)\int_0^1
p_1^{A_1-1}(1-p_1)^{B_1-1}\,dp_1
- \int_0^1
p_1^{A_1-1}(1-p_1)^{B_1-1}B_{p_1}(A_2,B_2)\,dp_1 \\
&= B(A_2,B_2)B(A_1,B_1) - B(A_2,B_2)\,J,
\end{aligned}
\]
where
\[
J
= \int_0^1
p_1^{A_1-1}(1-p_1)^{B_1-1}I_{p_1}(A_2,B_2)\,dp_1.
\]
Thus
\[
I(y_1,y_2)
= B(A_2,B_2)\bigl[B(A_1,B_1) - J\bigr].
\]
which can be substituted into \Cref{eq:predDensH+}.
\end{proof}

\begin{proof}[Finite-sum form for $I(y_1,y_2)$ (integer shapes)]
When \(A_2\) is a positive integer, we can again expand
\[
I_{p_1}(A_2,B_2)
= \frac{1}{B(A_2,B_2)}
\sum_{k=0}^{A_2-1}
\binom{A_2+B_2-1}{k}\,
p_1^{k}(1-p_1)^{A_2+B_2-1-k},
\]
to obtain
\[
\begin{aligned}
J
&= \int_0^1
p_1^{A_1-1}(1-p_1)^{B_1-1}I_{p_1}(A_2,B_2)\,dp_1 \\
&= \frac{1}{B(A_2,B_2)}
\sum_{k=0}^{A_2-1}
\binom{A_2+B_2-1}{k}
\int_0^1
p_1^{A_1-1+k}(1-p_1)^{B_1-1+A_2+B_2-1-k}\,dp_1 \\
&= \frac{1}{B(A_2,B_2)}
\sum_{k=0}^{A_2-1}
\binom{A_2+B_2-1}{k}
B\bigl(A_1+k,\;B_1+A_2+B_2-1-k\bigr).
\end{aligned}
\]
Therefore
\[
I(y_1,y_2)
= B(A_2,B_2)B(A_1,B_1)
- \sum_{k=0}^{A_2-1}
\binom{A_2+B_2-1}{k}
B\bigl(A_1+k,\;B_1+A_2+B_2-1-k\bigr).
\]
Plugging this into \Cref{eq:predDensH+}, that is, into
\[
p(y_1,y_2 \mid H_+)
= \binom{n_1}{y_1}\binom{n_2}{y_2}
\frac{I(y_1,y_2)}{B(a_1^d,b_1^d)B(a_2^d,b_2^d)\,C},
\]
and using the finite-sum form for \(C\) in \Cref{eq:finiteSumFormC}, yields an explicit expression
for the predictive density under \(H_+\) entirely in terms of Beta functions.
\end{proof}

\section{Derivations for the two-sided hypothesis test of $H_0:\eta = 0$ versus $H_-:\eta < 0$}

\subsection{One-sided hypothesis test of $H_0:\eta=0$ versus $H_-:\eta<0$}

\subsection*{Binomial model and hypotheses}

We observe
\[
Y_1 \sim \mathrm{Bin}(n_1,p_1),
\qquad
Y_2 \sim \mathrm{Bin}(n_2,p_2),
\]
and wish to test
\[
H_0 : \eta = 0 \quad\text{vs.}\quad H_- : \eta < 0,
\qquad
\eta = p_2 - p_1.
\]
Under \(H_0\) we have \(p_1 = p_2 =: p\).
Under \(H_-\) we impose the order constraint \(p_2 < p_1\).

\subsection{Design and analysis priors under $H_0$ and $H_-$}

The priors under $H_0$ remain identical to the previous section:
\[
p \sim \mathrm{Beta}(a_0^d,b_0^d) \quad\text{(design)}, \qquad
p \sim \mathrm{Beta}(a_0^a,b_0^a) \quad\text{(analysis)},
\]
with prior density
\[
\pi_0(p)
= \frac{1}{B(a_0^d,b_0^d)}\,p^{a_0^d-1}(1-p)^{b_0^d-1},
\qquad 0<p<1.
\]

Under $H_-$ we start from independent Beta design priors
\[
p_1 \sim \mathrm{Beta}(a_1^d,b_1^d),
\qquad
p_2 \sim \mathrm{Beta}(a_2^d,b_2^d),
\]
with joint (untruncated) density
\[
\pi_{\text{untr}}(p_1,p_2)
= \frac{1}{B(a_1^d,b_1^d)B(a_2^d,b_2^d)}\,
p_1^{a_1^d-1}(1-p_1)^{b_1^d-1}\,
p_2^{a_2^d-1}(1-p_2)^{b_2^d-1},
\qquad 0<p_1,p_2<1.
\]
We then truncate to the order-restricted region
\[
A_- = \{(p_1,p_2)\colon 0 < p_2 < p_1 < 1\}.
\]
The normalizing constant for the truncated prior is
\[
C_-
= \iint_{A_-} \pi_{\text{untr}}(p_1,p_2)\,dp_1\,dp_2
= P(p_2<p_1)
\]
under the independent Beta priors. A convenient expression is
\begin{align}\label{eq:Cminus}
C_-
= \int_0^1
\frac{p_1^{a_1^d-1}(1-p_1)^{b_1^d-1}}{B(a_1^d,b_1^d)}\,
I_{p_1}(a_2^d,b_2^d)\,dp_1.
\end{align}
Hence the truncated prior density under $H_-$ is
\[
\pi_-(p_1,p_2)
= \frac{\pi_{\text{untr}}(p_1,p_2)\,\mathbf{1}\{0<p_2<p_1<1\}}{C_-}.
\]

\subsection{Finite-sum form for $C_-$ (integer shapes)}

When $a_2^d$ is a positive integer,
\[
I_{p_1}(a_2^d,b_2^d)
= \frac{1}{B(a_2^d,b_2^d)}
\sum_{k=0}^{a_2^d-1}
\binom{a_2^d+b_2^d-1}{k}\,
p_1^{k}(1-p_1)^{a_2^d+b_2^d-1-k},
\]
yielding
\begin{align}\label{eq:finiteSumFormCminus}
C_-
&= \frac{1}{B(a_1^d,b_1^d)B(a_2^d,b_2^d)}
\sum_{k=0}^{a_2^d-1}
\binom{a_2^d+b_2^d-1}{k}\,
B\bigl(a_1^d+k,\;b_1^d+a_2^d+b_2^d-1-k\bigr).
\end{align}

\subsection{Predictive density under $H_0$}

The predictive density under $H_0$ is unchanged:
\begin{align}\label{eq:predDensH0minus}
p(y_1,y_2 \mid H_0)
= \binom{n_1}{y_1}\binom{n_2}{y_2}
\frac{B\bigl(y_1+y_2+a_0^d,\;n_1+n_2-y_1-y_2+b_0^d\bigr)}{B(a_0^d,b_0^d)}.
\end{align}

\subsection{Predictive density under $H_-$}

The likelihood remains
\[
f(y_1,y_2 \mid p_1,p_2)
= \binom{n_1}{y_1}\binom{n_2}{y_2}
p_1^{y_1}(1-p_1)^{n_1-y_1}
p_2^{y_2}(1-p_2)^{n_2-y_2}.
\]
The predictive density under $H_-$ is
\[
p(y_1,y_2 \mid H_-)
= \iint_{A_-} f(y_1,y_2 \mid p_1,p_2)\pi_-(p_1,p_2)\,dp_1\,dp_2.
\]
Substituting the truncated prior gives
\[
\begin{aligned}
p(y_1,y_2 \mid H_-)
&= \binom{n_1}{y_1}\binom{n_2}{y_2}
\frac{1}{B(a_1^d,b_1^d)B(a_2^d,b_2^d)\,C_-}
\int_0^1\int_0^{p_1}
p_2^{y_2+a_2^d-1}(1-p_2)^{n_2-y_2+b_2^d-1} \\
&\qquad\qquad\qquad\qquad\quad
\times p_1^{y_1+a_1^d-1}(1-p_1)^{n_1-y_1+b_1^d-1}
\,dp_2\,dp_1.
\end{aligned}
\]
Using the updated shape parameters from \Cref{eq:updatedShapeParameters},
the inner integral is
\[
\int_0^{p_1} p_2^{A_2-1}(1-p_2)^{B_2-1}\,dp_2
= B_{p_1}(A_2,B_2)
= B(A_2,B_2)\,I_{p_1}(A_2,B_2).
\]
Hence
\[
\begin{aligned}
p(y_1,y_2 \mid H_-)
&= \binom{n_1}{y_1}\binom{n_2}{y_2}
\frac{B(A_2,B_2)}{B(a_1^d,b_1^d)B(a_2^d,b_2^d)\,C_-}
\int_0^1
p_1^{A_1-1}(1-p_1)^{B_1-1}
I_{p_1}(A_2,B_2)
\,dp_1.
\end{aligned}
\]
Define
\[
J(y_1,y_2)
= \int_0^1
p_1^{A_1-1}(1-p_1)^{B_1-1}I_{p_1}(A_2,B_2)\,dp_1,
\]
so
\begin{align}\label{eq:predDensHminus}
p(y_1,y_2 \mid H_-)
= \binom{n_1}{y_1}\binom{n_2}{y_2}
\frac{B(A_2,B_2)J(y_1,y_2)}{B(a_1^d,b_1^d)B(a_2^d,b_2^d)\,C_-}.
\end{align}

\subsection{Finite-sum form for $J(y_1,y_2)$ (integer shapes)}

When $A_2$ is a positive integer,
\[
J(y_1,y_2)
= \frac{1}{B(A_2,B_2)}
\sum_{k=0}^{A_2-1}
\binom{A_2+B_2-1}{k}
B\bigl(A_1+k,\;B_1+A_2+B_2-1-k\bigr).
\]

\subsection{Bayes factor for $H_0:\eta=0$ versus $H_-:\eta<0$}

The Bayes factor for $H_-$ versus $H_0$ is
\[
\mathrm{BF}_{-0}(y_1,y_2)
= \frac{p(y_1,y_2 \mid H_-)}{p(y_1,y_2 \mid H_0)},
\]
with expressions from \Cref{eq:predDensH0minus,eq:predDensHminus}:
\begin{align}\label{eq:BF-0}
\mathrm{BF}_{-0}(y_1,y_2)
&= \frac{B(a_0^a,b_0^a)B(A_2,B_2)J(y_1,y_2)}{B(y_1+y_2+a_0^a,\;n_1+n_2-y_1-y_2+b_0^a)B(a_1^a,b_1^a)B(a_2^a,b_2^a)C_-},
\end{align}
where $C_-$ uses \Cref{eq:Cminus} or \Cref{eq:finiteSumFormCminus} for integer shapes, and $J(y_1,y_2)$ uses the finite-sum form above. When shape parameters are integers, the Bayes factor $\mathrm{BF}_{-0}$ can be computed entirely using Beta functions without numerical integration. Note that in \Cref{eq:BF-0}, we have replaced the design prior hyperparameters $a_i^d$ and $b_i^d$ of \Cref{eq:predDensH0minus,eq:predDensHminus} by analysis prior hyperparameters $a_i^a$, $b_i^a$, $i=1,2$.

\section{Derivations for the two-sided hypothesis test of $H_-:\eta \leq 0$ versus $H_+:\eta > 0$}

\subsection{Binomial model and hypotheses}

We observe
\[
Y_1 \sim \mathrm{Bin}(n_1,p_1),
\qquad
Y_2 \sim \mathrm{Bin}(n_2,p_2),
\]
and wish to test
\[
H_0 : \eta \leq 0 \quad\text{vs.}\quad H_1 : \eta > 0,
\qquad
\eta = p_2 - p_1.
\]
Under $H_0$ we impose the order constraint $p_2 \leq p_1$. Under $H_1$ we impose $p_2 > p_1$.

\subsection{Design and analysis priors under $H_0$ and $H_1$}

Under $H_0:p_2 \leq p_1$ we start from independent Beta design priors
\[
p_1 \sim \mathrm{Beta}(a_1^d,b_1^d),
\qquad
p_2 \sim \mathrm{Beta}(a_2^d,b_2^d),
\]
with joint (untruncated) density
\[
\pi_{\text{untr}}(p_1,p_2)
= \frac{1}{B(a_1^d,b_1^d)B(a_2^d,b_2^d)}\,
p_1^{a_1^d-1}(1-p_1)^{b_1^d-1}\,
p_2^{a_2^d-1}(1-p_2)^{b_2^d-1},
\qquad 0<p_1,p_2<1.
\]
We truncate to the order-restricted region
\[
A_0 = \{(p_1,p_2)\colon 0 \leq p_2 \leq p_1 \leq 1\}.
\]
The normalizing constant for the truncated prior under $H_0$ is
\[
C_0
= \iint_{A_0} \pi_{\text{untr}}(p_1,p_2)\,dp_1\,dp_2
= P(p_2 \leq p_1)
= 1 - P(p_2 > p_1).
\]
Using the expression for $P(p_2>p_1)$ from \Cref{eq:C},
\begin{align}\label{eq:C0}
C_0 = 1 - C,
\qquad
C = \int_0^1
\frac{p_2^{a_2^d-1}(1-p_2)^{b_2^d-1}}{B(a_2^d,b_2^d)}\,
I_{p_2}(a_1^d,b_1^d)\,dp_2.
\end{align}
The truncated prior density under $H_0$ is
\[
\pi_0(p_1,p_2)
= \frac{\pi_{\text{untr}}(p_1,p_2)\,\mathbf{1}\{0\leq p_2 \leq p_1 \leq 1\}}{C_0}.
\]

Under $H_1:p_2 > p_1$ the priors are identical to the $H_+$ case from the previous section:
\[
\pi_+(p_1,p_2)
= \frac{\pi_{\text{untr}}(p_1,p_2)\,\mathbf{1}\{0<p_1<p_2<1\}}{C},
\]
with normalizing constant $C$ from \Cref{eq:C}.

\subsection{Finite-sum form for $C_0$ (integer shapes)}

Using the finite-sum form for $C$ from \Cref{eq:finiteSumFormC}, we arrive at
\begin{align}\label{eq:finiteSumFormC0}
C_0 = 1 - \frac{1}{B(a_1^d,b_1^d)B(a_2^d,b_2^d)}
\sum_{k=0}^{a_1^d-1}
\binom{a_1^d+b_1^d-1}{k}\,
B\bigl(a_2^d+k,\;b_2^d+a_1^d+b_1^d-1-k\bigr).
\end{align}

\subsection{Predictive density under $H_0:\eta\leq 0$}

The predictive density under $H_0:\eta \leq 0$ can be derived as follows. 
The likelihood is
\[
f(y_1,y_2 \mid p_1,p_2)
= \binom{n_1}{y_1}\binom{n_2}{y_2}
p_1^{y_1}(1-p_1)^{n_1-y_1}
p_2^{y_2}(1-p_2)^{n_2-y_2}.
\]
The predictive density under $H_0$ is
\[
p(y_1,y_2 \mid H_0)
= \iint_{A_0} f(y_1,y_2 \mid p_1,p_2)\pi_0(p_1,p_2)\,dp_1\,dp_2.
\]
Substituting the truncated prior gives
\[
\begin{aligned}
p(y_1,y_2 \mid H_0)
&= \binom{n_1}{y_1}\binom{n_2}{y_2}
\frac{1}{B(a_1^d,b_1^d)B(a_2^d,b_2^d)\,C_0}
\int_0^1\int_0^{p_1}
p_2^{y_2+a_2^d-1}(1-p_2)^{n_2-y_2+b_2^d-1} \\
&\qquad\qquad\qquad\qquad\quad
\times p_1^{y_1+a_1^d-1}(1-p_1)^{n_1-y_1+b_1^d-1}
\,dp_2\,dp_1.
\end{aligned}
\]
Using updated shape parameters from \Cref{eq:updatedShapeParameters}, the inner integral is
\[
\int_0^{p_1} p_2^{A_2-1}(1-p_2)^{B_2-1}\,dp_2
= B_{p_1}(A_2,B_2)
= B(A_2,B_2)\,I_{p_1}(A_2,B_2).
\]
Hence
\[
\begin{aligned}
p(y_1,y_2 \mid H_0)
&= \binom{n_1}{y_1}\binom{n_2}{y_2}
\frac{B(A_2,B_2)}{B(a_1^d,b_1^d)B(a_2^d,b_2^d)\,C_0}
\int_0^1
p_1^{A_1-1}(1-p_1)^{B_1-1}
I_{p_1}(A_2,B_2)
\,dp_1.
\end{aligned}
\]
Define
\[
J(y_1,y_2)
= \int_0^1
p_1^{A_1-1}(1-p_1)^{B_1-1}I_{p_1}(A_2,B_2)\,dp_1,
\]
so
\begin{align}\label{eq:predDensH0composite}
p(y_1,y_2 \mid H_0)
= \binom{n_1}{y_1}\binom{n_2}{y_2}
\frac{B(A_2,B_2)J(y_1,y_2)}{B(a_1^d,b_1^d)B(a_2^d,b_2^d)\,C_0}.
\end{align}

\subsection{Finite-sum form for $J(y_1,y_2)$ (integer shapes)}

When $A_2$ is a positive integer, we arrive at
\[
J(y_1,y_2)
= \frac{1}{B(A_2,B_2)}
\sum_{k=0}^{A_2-1}
\binom{A_2+B_2-1}{k}
B\bigl(A_1+k,\;B_1+A_2+B_2-1-k\bigr).
\]

\subsection{Predictive density under $H_1:\eta>0$}

The predictive density under $H_1$ is identical to $H_+$ from the previous section. For the hypothesis $H_+ : p_2 > p_1 $, the joint predictive probability $ f(y_1, y_2 | H_+) $ can be derived similarly to the $ H_- $ case but with the integration limit reflecting the inequality $ p_2 > p_1 $. We have

$$ p_1 \sim \mathrm{Beta}(a_1^d, b_1^d), \hspace{1cm} y_1 \sim \mathrm{Binomial}(n_1, p_1)$$
$$ p_2 \sim \mathrm{Beta}(a_2^d, b_2^d), \hspace{1cm} y_2 \sim \mathrm{Binomial}(n_2, p_2)$$
The joint prior-predictive probability under $ H_+ $ can be written as
$$
\begin{aligned}
f(y_1, y_2 | H_+) &= \iint_{0 \leq p_1 \leq 1, \, p_1 \leq p_2 \leq 1} f(y_1|p_1) f(p_1|a_1^d,b_1^d) 
\times f(y_2|p_2) f(p_2|a_2^d,b_2^d) \, dp_2 \, dp_1,
\end{aligned}
$$
where
$$
f(y_i | p_i) = \binom{n_i}{y_i} p_i^{y_i} (1-p_i)^{n_i - y_i},
$$
and
$$
f(p_i | a_i^d, b_i^d) = \frac{p_i^{a_i^d -1}(1-p_i)^{b_i^d -1}}{\mathrm{B}(a_i^d, b_i^d)}.
$$
Now, we can simplify the inner integral over $p_2$ as follows:
$$
\begin{aligned}
\int_{p_1}^1 f(y_2|p_2) f(p_2|a_2^d, b_2^d) \, dp_2 &= \binom{n_2}{y_2} \frac{1}{\mathrm{B}(a_2^d, b_2^d)} \int_{p_1}^1 p_2^{y_2 + a_2^d - 1} (1 - p_2)^{n_2 - y_2 + b_2^d -1} \, dp_2 \\
&= \binom{n_2}{y_2} \frac{\mathrm{B}_{1} (y_2 + a_2^d, n_2 - y_2 + b_2^d) - \mathrm{B}_{p_1} (y_2 + a_2^d, n_2 - y_2 + b_2^d)}{B(a_2^d, b_2^d)},
\end{aligned}
$$
where $ \mathrm{B}_x(a,b) $ is the incomplete Beta function and $ \mathrm{B}_1(a,b) = \mathrm{B}(a,b) $. The integral simplifies as:
$$
\int_{p_1}^1 \text{Beta PDF}(p_2; a', b') \, dp_2 = 1 - I_{p_1}(a', b'),
$$
where $ I_x(a,b) $ is the regularized incomplete Beta function. Thus, the final expression for $ f(y_1, y_2 | H_+) $ is:
\begin{align}\label{eq:predDensH1composite}
f(y_1, y_2 | H_+) &= \binom{n_1}{y_1} \binom{n_2}{y_2} \frac{1}{\mathrm{B}(a_1^d, b_1^d) \mathrm{B}(a_2^d, b_2^d)} \nonumber\\
&\times \int_0^1 p_1^{y_1 + a_1^d - 1} (1-p_1)^{n_1 - y_1 + b_1^d - 1} \left[1 - I_{p_1}(y_2 + a_2^d, n_2 - y_2 + b_2^d)\right] dp_1.
\end{align}

\subsection{Bayes factor for $H_-:\eta\leq 0$ versus $H_+:\eta>0$}

The Bayes factor for $H_1$ ($H_+$) versus $H_0$ ($H_-$) is
\[
\mathrm{BF}_{10}(y_1,y_2)
= \frac{p(y_1,y_2 \mid H_1)}{p(y_1,y_2 \mid H_0)},
\]
with expressions from \Cref{eq:predDensH0composite,eq:predDensH1composite}:
\begin{align}\label{eq:BF10composite}
\mathrm{BF}_{10}(y_1,y_2)
&= \frac{\frac{I(y_1,y_2)}{B(a_1^a,b_1^a)B(a_2^a,b_2^a)\,C}}{\frac{B(A_2,B_2)J(y_1,y_2)}{B(a_1^a,b_1^a)B(a_2^a,b_2^a)\,C_0}}
= \frac{I(y_1,y_2)C_0}{B(A_2,B_2)J(y_1,y_2)C},
\end{align}
where $C_0=1-C$ uses \Cref{eq:C0,eq:finiteSumFormC0}, $I(y_1,y_2)$ uses the $H_+$ finite-sum form, and $J(y_1,y_2)$ uses the finite-sum above in case $C_0$ or $A_2$ are (positive) integers. When shape parameters are integers, $\mathrm{BF}_{10}$ can be therefore computed entirely using Beta functions. Note that in \Cref{eq:BF10composite} we have replaced the design prior hyperparameters $a_i^d$ and $b_i^d$ for $i=1,2$ by analysis prior hyperparameters $a_i^a$ and $b_i^a$, $i=1,2$.

\bibliography{library} 

\end{document}